\documentclass[11pt,a4paper]{emulateapj}
\usepackage{epsfig}
\usepackage{amsmath}
\usepackage{psfig}
\usepackage{graphicx}
\usepackage{rotating}
\usepackage{longtable}

\usepackage{natbib}

\def\apj{ApJ}
\def\apjl{ApJL}
\def\mnras{MNRAS}
\def\pasp{PASP}

\def\araa{ARAA}
\def\aap{A\&A}
\def\aj{AJ}
\def\apjs{ApJS}

\def\nat{Nature}

\def\gs{\mathrel{\raise0.35ex\hbox{$\scriptstyle >$}\kern-0.6em\lower0.40ex\hbox{{$\scriptstyle \sim$}}}} 
\def\ls{\mathrel{\raise0.35ex\hbox{$\scriptstyle <$}\kern-0.6em\lower0.40ex\hbox{{$\scriptstyle \sim$}}}}

\def\Wm2{\,\hbox{W}\,\hbox{m}^{-2}} 
\def\gsim{\mathrel{\raise0.35ex\hbox{$\scriptstyle >$}\kern-0.6em\lower0.40ex\hbox{{$\scriptstyle \sim$}}}} 
\def\lsim{\mathrel{\raise0.35ex\hbox{$\scriptstyle <$}\kern-0.6em\lower0.40ex\hbox{{$\scriptstyle \sim$}}}} 
\def\ltsima{$\; \buildrel < \over \sim \;$} 
\def\simlt{\lower.5ex\hbox{\ltsima}} 
\def\gtsima{$\; \buildrel > \over \sim \;$} 
\def\simgt{\lower.5ex\hbox{\gtsima}}

\lefthead{Danielson et al.}  \righthead{A spectroscopic redshift
  survey of ALMA-identified submillimetre galaxies}

\begin{document}

\title{An ALMA survey of submillimetre galaxies in the Extended {\it Chandra} Deep Field South: Spectroscopic redshifts}

\author{
A.\,L.\,R.\ Danielson,\altaffilmark{1}
A.\,M.\ Swinbank,\altaffilmark{1,2,*}
Ian Smail,\altaffilmark{1,2}
J.\,M.\ Simpson,\altaffilmark{1}
C.\,M.\,Casey,\altaffilmark{3,4}
S.\,C.\,Chapman,\altaffilmark{5}
E.\ da Cunha,\altaffilmark{6}
J.\,A.\ Hodge,\altaffilmark{7}
F.\ Walter,\altaffilmark{8}
J.\,L.\ Wardlow,\altaffilmark{1}
D.\,M.\ Alexander,\altaffilmark{1}
W.\,N.\ Brandt,\altaffilmark{9}
C.\ de Breuck,\altaffilmark{10}
K.\,E.\,K.\ Coppin\altaffilmark{11}
H. Dannerbauer,\altaffilmark{12}
M.\ Dickinson,\altaffilmark{13}
A.\,C.\ Edge,\altaffilmark{1}
E.\ Gawiser,\altaffilmark{14}
R.\,J.\ Ivison,\altaffilmark{10}
A.\ Karim,\altaffilmark{17}
A.\ Kovacs,\altaffilmark{15}
D.\ Lutz,\altaffilmark{16}
K.\ Menten,\altaffilmark{17}
E.\, Schinnerer,\altaffilmark{7}
A.\ Wei{\ss},\altaffilmark{17}
P.\ van der Werf,\altaffilmark{7}
}
\altaffiltext{1}{Centre for Extragalactic Astronomy, Durham University, Department of Physics, South Road, Durham DH1 3LE, UK}
\altaffiltext{2}{Institute for Computational Cosmology, Durham University, South Road, Durham DH1 3LE, UK}
\altaffiltext{*}{Email: a.m.swinbank@durham.ac.uk}
\altaffiltext{3}{Department of Astronomy, The University of Texas at Austin, 2515 Speedway Boulevard Stop C1400, Austin, TX 78712, USA}
\altaffiltext{4}{Department of Physics and Astronomy, University of California, Irvine, Irvine, CA 92697, USA}
\altaffiltext{5}{Department of Physics and Atmospheric Science, Dalhousie University, Halifax, NS B3H 4R2, Canada}
\altaffiltext{6}{Research School of Astronomy and Astrophysics, Australian National University, Canberra, ACT 2611, Australia}
\altaffiltext{7}{Leiden Observatory, Leiden University, P.O. Box 9513, 2300 RA Leiden, The Netherlands 0000-0001-5434-5942}
\altaffiltext{8}{Max-Planck-Institut f\"ur Astronomie, K\"onigstuhl 17, D-69117 Heidelberg, Germany}
\altaffiltext{9}{Department of Astronomy and Astrophysics and the Institute for Gravitation and the Cosmos, The Pennsylvania State University}
\altaffiltext{10}{European Southern Observatory, Karl Schwarzschild Stra{\ss}e 2, 85748, Garching, Germany}
\altaffiltext{11}{Centre for Astrophysics Research, Science and Technology Research Institute, University of Hertfordshire, College Lane, Hatfield AL10 9AB, UK }
\altaffiltext{12}{Universit\"at Wien, Institut f\"ur Astrophysik, T\"urkenschanzstra{\ss}e 17, 1180, Wien, Austria}
\altaffiltext{13}{National Optical Astronomy Observatory, Tucson, AZ 85719, USA}
\altaffiltext{14}{Department of Physics and Astronomy, Rutgers University, Piscataway, NJ 08854-8019, USA}
\altaffiltext{15}{Astronomy Department, University of Minnesota, MN 12345, USA}
\altaffiltext{16}{Max-Planck-Institut f\"ur extraterrestrische Physik, Giessenbachstra{\ss}e, 85748, Garching, Germany}
\altaffiltext{17}{Max-Planck-Institut f\"ur Radioastronomie, Auf dem H\"ugel 69, D-53121 Bonn, Germany}

\begin{abstract}
  We present spectroscopic redshifts of S$_{870\mu \rm
    m}\gsim$\,2\,mJy submillimetre galaxies (SMGs) which have been
  identified from the ALMA follow-up observations of 870-$\mu$m
  detected sources in the Extended {\it Chandra} Deep Field South (the
  ALMA-LESS survey).  We derive spectroscopic redshifts for 52 SMGs,
  with a median of $z$\,=\,2.4\,$\pm$\,0.1.  However, the distribution
  features a high redshift tail, with $\sim$\,23\% of the SMGs at
  $z\geq$\,3.  Spectral diagnostics suggest that the SMGs are young
  starbursts, and the velocity offsets between the nebular emission
  and UV ISM absorption lines suggest that many are driving winds,
  with velocity offsets up to 2000\,km\,s$^{-1}$.  Using the
  spectroscopic redshifts and the extensive UV-to-radio photometry in
  this field, we produce optimised spectral energy distributions
  (SEDs) using {\sc Magphys}, and use the SEDs to infer a median
  stellar mass of
  $M_\star$\,=\,(6\,$\pm$\,1)\,$\times$\,10$^{10}$\,M$_{\odot}$ for
  our SMGs with spectroscopic redshift.  By combining these stellar
  masses with the star-formation rates (measured from the far-infrared
  SEDs), we show that SMGs (on average) lie a factor $\sim$\,5 above
  the so-called ``main-sequence'' at $z\sim$\,2.  We provide this
  library of 52 template fits with robust and uniquely well-sampled
  SEDs available as a resource for future studies of SMGs, and also
  release the spectroscopic catalog of $\sim$\,2000 (mostly
  infrared-selected) galaxies targeted as part of the spectroscopic
  campaign.
\end{abstract}


\section{Introduction}
\label{sec:intro}

Submillimeter galaxies (SMGs) with 850\,$\mu$m fluxes of S$_{\rm
  850}>$\,1\,mJy represent a population of dusty starbursts whose space
density peaked $\sim$\,10\,Gyr ago.  Although they are relatively
rare, their far-infrared luminosities (L$_{\rm
  IR}>$\,2\,$\times$\,10$^{12}$\,L$_\odot$) imply high star-formation
rates ($\gsim$\,300\,M$_\odot$\,yr$^{-1}$) and so SMGs appear to
contribute at least 20\% of the total cosmic star-formation rate density
over $z$\,=\,1--4
\citep[e.g.][]{Chapman05,Barger12,Casey14,Swinbank14}.  If they can
maintain their star-formation rates, SMGs also have the potential to
consume all their cold gas reservoir within just 100\,Myr
\citep[e.g.][]{Tacconi08,Bothwell13}, and so double their stellar
masses within their short but intense lifetime
\citep[e.g.][]{Hainline09,Magnelli12}.  Their ability to form up to
10$^{11}$\,M$_\odot$ of stars within a short period of time makes them
candidates of progenitors of $z$\,=\,1--2 compact quiescent
galaxies \citep{Toft14,Simpson15a,Ikarashi15} as well as local massive
ellipticals \citep[e.g.][]{Lilly99,Genzel03,Simpson14}.
These characteristics suggest that bright SMGs represent an essential
population for models of galaxy formation and evolution
\citep[e.g.][]{Efstathiou03,Baugh05,Swinbank08,Narayanan09,Dave10,Hayward11,Lacey16}.

However, to identify the physical processes that trigger the
starbursts, measure the internal dynamics of the cold (molecular) and
ionised gas, and infer stellar masses first requires accurate
redshifts.  
To date, the largest such spectroscopic survey of 870\,$\mu$m-selected submillimetre
sources was
carried out by \citet{Chapman05} who targeted a sample of 104
radio-identified, SCUBA-detected submillimetre sources spread across
seven extragalactic survey fields.  Using rest-frame UV
spectroscopy with the Low-resolution Imaging Spectrograph (LRIS) on
the Keck telescope, they derived spectroscopic redshifts for 73 submillimetre
sources with a median
redshift of $z\sim$\,2.4 for the radio-selected sample (with a 
maximum redshift in their sample of $z$\,=\,3.6).

Although the requirement for a radio detection in these previous
surveys was a necessary step to identify the most probable galaxy
counterpart responsible for the submillimetre emission, the radio wavelengths
do not benefit from the same negative K-correction as longer submillimetre
wavelengths and indeed, above $z\sim$\,3.5, the 1.4\,GHz flux of a
galaxy with a star-formation rate of
$\sim$\,100\,M$_{\odot}$\,yr$^{-1}$ falls below $\sim$\,15\,$\mu$Jy and
so below the typical sensitivity limit of deep radio surveys.  This
has the potential to bias the redshift distribution to $z\lsim$\,3.5,
especially if a significant fraction of submillimetre sources do not have
multi-wavelength counterparts.  Indeed, in single-dish 850\,$\mu$m
surveys, up to 50\% of all submillimetre sources are undetected at
radio wavelengths \citep[e.g.][]{Ivison05,Ivison07,Biggs11}.  Some
progress can be made by targeting lensed sources whose
multi-wavelength identifications are less ambiguous, and indeed
spectroscopic redshifts have been derived for SMGs up to $z\sim$\,5
\citep[e.g.][]{Weiss13}.

Due to the angular resolution and sensitivity of the ALMA
interferometer, it has become possible to identify the counterparts of
submillimetre sources to $\lsim$\,0.3$''$ accuracy without recourse to
statistical asssociations at other wavelengths.  To identify a sample
of SMGs in a well studied field with a well defined selection
function, \citet{Hodge13} undertook an ALMA survey of 122 SMGs found
in the Extended {\it Chandra} Deep Field South (ECDFS): the ``ALESS''
survey.  This survey followed up 122 of the 126 submillimetre sources
originally detected with the LABOCA instrument on the Atacama
Pathfinder Experiment 12 metre telescope (APEX); the LABOCA ECDFS
Sub-mm Survey (LESS) \citep{Weiss09}.  Each LESS submillimetre source
was targeted with ALMA at 870\,$\mu$m (Band 7).  The typical FWHM of
the ALMA synthesised beam was $\sim$\,1.5$''$ (significantly smaller
than the LABOCA 19.2$''$ beam), thus allowing us to directly pinpoint
the position of the SMG precisely.

From these data, \citet{Karim13} \citep[see also][]{Simpson15b} showed
that statistical identifications (e.g.\ using radio counterparts)
were incorrect in $\sim$\,30\% of cases, whilst the
single-dish submillimetre sources also suffer from significant
``multiplicity'', with $>$\,35\% of the single-dish sources resolved
into multiple SMGs brighter than $\gsim$\,1\,mJy.  This flux limit
corresponds approximately to a far-infrared luminosity of $L_{\rm
  FIR}\gsim$\,10$^{12}$\,L$_{\odot}$ at $z\sim$\,2, and so it appears
that a large fraction of the single-dish submillimetre sources often contain
two (or more) 
Ultra-Luminous Infrared Galaxies (ULIRGs).
Consequently, a new ALESS SMG catalogue was
defined comprising 131 SMGs \citep{Hodge13}.

One of the primary goals of the ALESS survey is to provide an unbiased catalog
of SMGs for which we can derive  molecular gas masses, as
well as measure spatially resolved dynamics of the gas and stars in
order to identify the trigering mechanisms that cause the burst of
star formation.  The first necessary step in this process is
to derive the precise spectroscopic redshifts.  To this end, we have
undertaken a spectroscopic survey of ALMA-identified SMGs using VLT,
Keck and Gemini (supplemented by ALMA) and in this paper we describe
the UV, optical and near-infrared spectroscopic follow-up.  We use the
resulting redshifts to investigate the redshift distribution, the
environments and typical spectral features of these SMGs.  In addition,
we use these precise redshifts to better constrain the SED fitting
from UV-to-radio wavelengths and provide template SEDs for the ALESS
SMG population.

The structure of the paper is as follows.  We discuss the observations
and the data reduction in \S~\ref{sec:obs2}, followed by redshift
identification and sample properties in \S~\ref{sec:anal2}.  In
\S~\ref{sec:specdist} we show the ALESS redshift distribution and
discuss the spectroscopic completeness.  In \S~\ref{sec:discuss2} we
discuss the velocity offsets of various different spectral lines,
search for evidence of stellar winds and galaxy-scale outflows and
investigate the environments of SMGs and the individual and composite
spectral properties.  We present our conclusions in \S~\ref{sec:conc}.
In the Appendix, we give the table of ALESS SMG redshifts and provide
information on individual SMGs from the sample.

Unless otherwise stated the quoted errors on the median values within
this work are determined through bootstrap analysis and are quoted as
the equivalent of 68.3\% confidence limits.  Throughout the
paper we use a $\Lambda$CDM cosmology with H$_{\rm
  0}$\,=\,72\,km\,s$^{-1}$\,Mpc$^{-1}$, $\Omega_{\rm m}$\,=\,0.27 and
$\Omega_{\rm \Lambda}$\,=\,1\,-\,$\Omega_{\rm m}$ \citep{Spergel04}
and a Chabrier initial mass function (IMF; \citealt{Chabrier03}).
Unless otherwise noted, all magnitudes are on the AB system.

\section{Observations and reduction}
\label{sec:obs2}

\subsection{Sample definition}
The 870\,$\mu$m LESS survey \citep{Weiss09} was undertaken using the
LABOCA camera on APEX, covering an area of 0.5$\times$0.5 degrees
centered on the ECDFS.  The total exposure time for the survey was
310\,hours, reaching a 1-$\sigma$ sensitivity of $\sigma_{870\mu \rm
  m}\sim$\,1.2\,mJy\,beam$^{-1}$ with a beam of 19.2$''$ FWHM.  In
total, we identified 126 submillimetre sources above a signal-to-noise
of 3.7$\sigma$.  Follow-up observations of the LESS sources were
carried out with ALMA (the ALMA-LESS, ALESS programme).  Details of the
ALMA observations are described in \cite{Hodge13} but in summary, the
120\,s observations for each source were taken between October and
November 2011 in the Cycle 0 Project \#2011.1.00294.S.  These
submillimetre interferometric identifications confirmed some of the
probabilistically determined counterparts \citep{Biggs11,Wardlow11}
but also revealed some mis-identified counterparts and a significant
number of new counterparts.  Therefore, the ALESS SMG catalogue was
formed, comprising a main (hereafter {\sc main}) catalogue of the 99 of
the most reliable ALMA-identified SMGs (i.e.\ lying within the the
primary beam FWHM of the best-quality maps).  A supplementary
(hereafter {\sc supp}) catalogue was also defined comprising 32
ALMA-identified SMGs extracted from outside the ALMA primary beam, or
in lower quality maps \citep{Hodge13}.  When searching for
spectroscopic redshifts, we included both the {\sc main} and {\sc
  supp} sources, and in \S~\ref{sec:specdist} we demonstrate that the
inclusion of {\sc supp} sources makes very little quantitative
difference to the statistics of the redshift distribution.

To search for spectroscopic redshifts, we initiated an observing
campaign using the the FOcal Reducer and low dispersion Spectrograph
(FORS2) and VIsible MultiObject Spectrograph (VIMOS) on VLT, but to
supplement these observations, and in particular to increase the
wavelength coverage and probability of determining redshifts, we also
obtained observations with XSHOOTER on VLT, the Gemini Near-Infrared
Spectrograph (GNIRS) and the Multi-Object Spectrometer for Infra-Red
Exploration (MOSFIRE) on the Keck {\sc i} telescope, all of which
cover the near-infrared.  As part of a spectroscopic campaign
targeting \emph{Herschel}-selected galaxies in the ECDFS, ALESS SMGs
were included on DEep Imaging Multi-Object Spectrograph (DEIMOS) slit
masks on Keck {\sc ii} \citep[e.g.][]{Casey12}.  These observations
probe a similar wavelength range to FORS2 targeting some of the
ALMA-identified SMGs that could not be targeted with VLT (due to
slit collisions).  In total, we observed 109 out of the 131 ALESS
SMGs in the combined {\sc main} and {\sc supp} samples.  
In many cases we have ALESS SMGs with spectra from five different
spectrographs covering a broad wavelength range and we can cross check the
spectroscopic redshifts across all of the instruments.  Next, we
discuss the various instruments involved in our survey.  We note that
for all observations described below, flux calibration was carried out
using standard stars to calibrate instrumental response.

\subsection{VLT FORS2\,/\,VIMOS}
\begin{figure}
\begin{center}{
\psfig{file=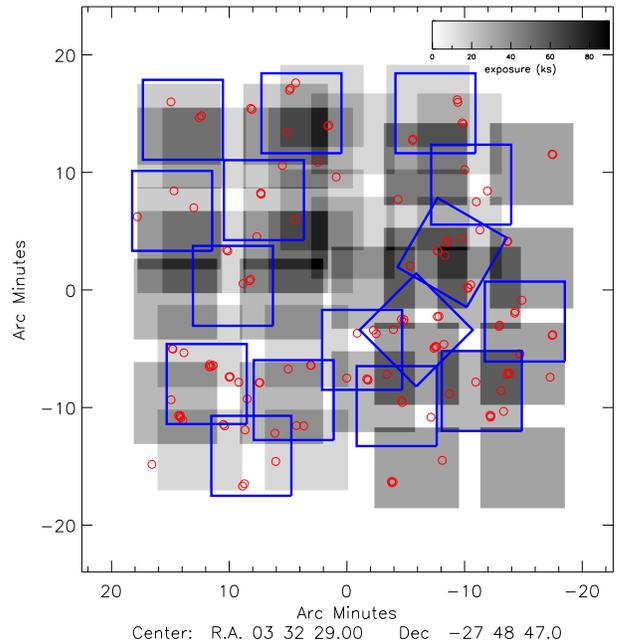,width=3.7in}
\caption{The coverage of our ten VIMOS pointings (greyscale) and 16
  FORS2 pointings (blue boxes) in the ECDFS. The ALESS SMG positions
  are shown as small red circles. VIMOS has four quadrants separated
  by small gaps. There is significant overlap between the VIMOS
  pointings, we therefore show the pointings here with the darkest
  areas corresponding to the regions with the longest total exposure
  time. Our FORS2/VIMOS programme covers 62 out of the 109
  targeted SMGs in the ECDFS. }
\label{fig:spec_coverf}
}\end{center}
\end{figure}

Our spectroscopic programme aimed to target as many of the ALESS SMGs
as possible using a dual approach with FORS2 and VIMOS (for a typical
SMG redshift of $z\sim$\,1--3, we are sensitive to Ly$\alpha$ and UV
ISM lines with VIMOS or [O{\sc ii}]\,$\lambda$3727 with FORS2).  In
total, we observed for 100\,hours each with VIMOS and FORS as part of
programme 183.A-0666.  We used deep exposures on ten (overlapping)
VIMOS masks to cover the field, plus deep integrations for sixteen
FORS masks (which cover a sub-set of the field but target the regions
with the highest density of ALMA SMGs; Fig.~\ref{fig:spec_coverf}).
All of the FORS observations were carried out in grey time and all of
the VIMOS observations carried out in dark time during service mode
runs with seeing $\le$\,0.8$''$ and clear sky conditions (transparency
variations below 10\%).  Our dual-instrument approach allowed us to
probe a large wavelength range using VIMOS LR-Blue grism
(4000--6700\AA\,) and FORS2 300I (6000--11000\,AA\,).  When designing
the slit masks, the first priority was always given to the SMGs, but
we also infilled the masks with other mid- or far-infrared selected
galaxies from the FIDEL {\it Spitzer} survey \citep{Magnelli09}, the
HerMES and PEP \textit{Herschel} surveys of this field
\citep{Oliver12,Lutz11}, $S_{1.4\rm GHz}>$\,30$\mu$\,Jy radio sources
and {\it Chandra} X-ray sources \citep{Lehmer05,Luo08} or
optical/near-infrared colour selected galaxies (see Table~3 and
Fig.~\ref{fig:NzALL}).

In Fig.~\ref{fig:spec_coverf} we show the spectroscopic coverage of
the ECDFS from our FORS2 and VIMOS programmes, where the darkest areas
demonstrate the areas with the longest total exposure time and the
FORS2 pointings are overlaid.  In total, we recorded 5221 galaxy
spectra, targeting 2454 (unique) galaxies.  

%
%

\subsubsection{FORS2}
FORS2 covers the the wavelength range $\lambda$\,=\,3300--11000{\AA}
and provides an image scale of 0.25$''$\,pix$^{-1}$ in the standard
readout mode (2$\times$2 binning). FORS2 was used in its multi-object
spectroscopy mode with exchangeable masks (MXU).  We varied the slit
length and orientation for each target in order to observe the maximum
number of sources on each mask (Fig.~\ref{fig:spec_coverf}), but we
consistently used a slit width of 1$''$.  We used $\sim$\,40--70
slits per mask and the OG590 order-sorting filter with the 300I grism
which results in a wavelength range covering
6000--11000{\AA}.  The typical resolution in this configuration is
$R=$\,$\lambda$\,/\,$\Delta\lambda\sim$\,660.  We used 16 pointings,
although in a small number of cases, we moved slits between exposures if
there were multiple sources within $\sim$\,5$''$ which could not be
simultaneously observed on a mask.  Each mask was observed in blocks
of 3\,$\times$\,900\,s with each exposure nodded up and down the slits
by $\sim$\,1.0$''$ to aid sky-subtraction and cosmic-ray removal when
the images were combined.  Each mask was typically observed six times
(with a range of three to nine times depending on the number of SMGs
on the mask and their median brightness), resulting in an on-source
exposure time 4.5\,hrs (with a range of 2.25--6.75\,hr).

We reduced the data using the spectroscopic reduction package from
\citet{Kelson03} adapted for use with FORS2 data FORS2
pipeline.  The pipeline produces two-dimensional, bias-corrected, flat-fielded,
wavelength-calibrated, sky-subtracted images.  Individual exposures
were combined in two-dimensions by taking a median of the frames and
sigma clipping.  We then extracted one-dimensional spectra over the
full spatial-extent of the continuum/emission lines visible, or in the
case where no emission was obvious in the two-dimensional image, we
extracted data from the region around the expected source position.

\subsubsection{VIMOS}
The VIMOS observations were undertaken in multi-object spectroscopy
(MOS) mode.  VIMOS consists of four quadrants each of a field-of-view
of 7$\,'\times$\,8$'$ with a detector pixel scale of 0.205$''$\,pix$^{-1}$.  Each
observing block comprised 3\,$\times$\,1200\,s exposures dithering
$\pm$\,1.0$''$ along the slit. The exposure time per mask was
3--9\,hr, again depending on the number of SMGs on the mask and
their average brightness.  Slit widths of 1.0$''$ were used, for which
the typical resolution is $R\sim$\,180 and the dispersion is
5.3\AA\,pix$^{-1}$ for the LR\_blue grism with the OS\_blue order sorting
filter ($\sim$\,4000--6700{\AA}).  We used 40--160 slits per
quadrant, totalling 160--400 slits over the four quadrants.  The
data were reduced using the standard {\sc ESOREX} pipeline package for
VIMOS.  The frames were stacked in two-dimensions before extracting
the one-dimensional spectra.  In a number of cases, the data suffer
from overlapping spectra which results in a second order overlapping
the adjacent spectrum (this can be seen in the VIMOS two-dimensional
spectrum of ALESS\,057.1 in Fig.~\ref{fig:spec1}).

\subsection{XSHOOTER}
To improve the wavelength coverage of our observations, we also
obtained XSHOOTER observations of 20 ALESS SMGs.  XSHOOTER
simultaneously observes from UV to near-infrared wavelengths covering
wavelength ranges of 3000--5600{\AA}, 5500--10200{\AA} and
10200--24800{\AA} for the UV (UVB), visible (VIS) and
near-infrared (NIR) arms respectively.  Targets were prioritised for
XSHOOTER follow-up based on their $K$-band magnitudes.  Our XSHOOTER
observations were taken in visitor mode as part of programme
090.A-0927(A) from 2012 December 7--10 in dark time.  We observed
each source for $\sim$\,1\,hr in generally clear conditions with a typical
seeing of $\sim$\,1.0$''$.  Our observing strategy was
4\,$\times$\,600\,s exposures per source, nodding the source up and
down the slit. The pixel scales were 0.16, 0.16 and
0.21$''$\,pix$^{-1}$ for the UVB, VIS and NIR arms respectively.  The
slits were all 11$''$ long and 0.9$''$ wide for the VIS and NIR arms
and 1.0$''$ wide for the UVB arm.  The typical resolution was
$R\sim$\,4350, 7450, 5300 for the UVB, VIS and NIR arms respectively.
The data reduction was carried out using the standard {\sc esorex}
pipeline package for XSHOOTER.

\subsection{MOSFIRE}
We also targeted 36 ALESS SMGs with the MOSFIRE spectrograph on Keck
{\sc i} (2012B\_H251M, 2013B\_U039M, and 2013B\_N114M) in $H$-
(1.46--1.81\,$\mu$m) and $K$-band (1.93--2.45\,$\mu$m).
Observations were taken in clear or photometric conditions with the
seeing varying from 0.4--0.9$''$.  In all cases we used slits of
width 0.7$''$.  The pixel scale of MOSFIRE is 0.18$''$\,pix$^{-1}$ and
the typical spectral resolution for this slit width is $R\sim$\,3270.
The total exposure time per mask was 2.2--3.6\,ks which was split
in to 120\,s ($H$-band) and 180\,s ($K$-band) exposures, with an ABBA
sequence and a 1.5$''$ nod along the slit between exposures.  Data
reduction was completed with {\sc mospy}.

\subsection{DEIMOS}
We targeted 71 of the ALESS SMGs as ``mask infill'' during a Keck {\sc
  ii} DEIMOS spectroscopy programme to measure redshifts for
\emph{Herschel}\,/\,SPIRE sources (programme 2012B\_H251).  The data
were taken on 2012 December 9--10 in clear conditions with seeing
between 1--1.3$''$.  We used a setup with the 600ZD (600 lines
mm$^{-1}$) grating with a 7200$\mbox{\AA}$ blaze angle and the GG455
blocking filter which resulted in a wavelength range of
4850--9550{\AA}.  Slit widths of 0.75$''$ were used and the masks
were filled with 40--70 slits per mask.  The pixel scale of DEIMOS
is 0.1185$''$\,pix$^{-1}$ and the typical resolution was $R\sim$\,3000.
Individual exposures were 1200\,s, and the total integration times
were 2--3\,hrs.  The data were reduced using the DEEP2 DEIMOS data
reduction pipeline \citep{Cooper12,Newman13}.

\subsection{GNIRS}
The Gemini Near-Infrared Spectrograph (GNIRS) was used to target eight
ALESS SMGs as (programme GN-2012B-Q-90) between 2012 November 10--15
and December 4--23.  The targets were selected based on their
$K$-band magnitude and whether they had a photometric redshift that
was predicted to place strong emission lines in the near-infrared.
The instrument was used in cross-dispersing mode (via the SXD prism
with 32 lines\,mm$^{-1}$), using the short camera, with slit widths of
0.3$''$, slit lengths of 7$''$ and a pixel scale of
0.15$''$\,pix$^{-1}$.  The wavelength coverage with this setup is
9000--25600{\AA}, typically with $R\sim$\,1700.  Our observing
strategy comprised 200\,s exposures and nodding up and down the slit
by $\sim$\,1$''$.  Each observing block comprised eight coadds of three
exposures, resulting in an exposure of $\sim$\,1.3\,hr per source. The
GNIRS data were reduced using the Gemini {\sc iraf} package.

\subsection{ALMA}
Spectroscopic redshifts for two of our SMGs, ALESS\,61.1 and
ALESS\,65.1 were determined from serendipitous detections of the
[C{\sc ii}]$\lambda$158$\mu$m line in the ALMA band
\citep{Swinbank12c}.  Although based on single line identifications,
both redshifts have been confirmed by the identification of
$^{12}$CO(1--0) emission using ATCA (\citealt{Huynh13}; Huynh et
al.\ 2017, submitted).
\medskip

\noindent Once all of the data were collected from the different spectrographs,
we collated the spectra for each ALESS SMG.  The instruments used to
observe each SMG are listed in Table~1.

\section{Analysis}
\label{sec:anal2}

\subsection{Redshift identification}
To determine redshifts for the sample, the one- and two-dimensional
spectra (for all $\sim$\,2000 galaxies) were independently examined by
two investigators (AMS and ALRD).  Any emission\,/\,absorption
features that were identified were fit with a Gaussian profiles to
determine their central wavelengths.  In the FORS2, VIMOS and DEIMOS
data the most commonly identified lines were Ly$\alpha$, C{\sc
  iv}\,$\lambda\lambda$1548.89,1550.77\,\AA, C{\sc
  iii}\,$\lambda$1909\,\AA, He{\sc ii}\,$\lambda$1640\,AA\ and [O{\sc
    ii}]$\lambda\lambda$3726.03,3728.82\,\AA.  In the near-infrared,
we typically detect H$\alpha$, N{\sc ii}\,$\lambda$6583 and [O{\sc
    iii}]\,$\lambda\lambda$4959, 5007 and in a small number of cases,
H$\beta$ (see Tables~2 \& 3).  The
optical\,/\,near-infrared counterparts of the SMGs are often faint and
we detect continuum in only $\sim$\,50\% of the 52 SMGs for which we
determine a redshift, (compared to $\sim$\,75\% for the
radio-identified submillimetre sources in \citealt{Chapman05}).

\setcounter{footnote}{0}

The spectra often only contain weak continuum, emission and\,/\,or
absorption lines, making redshifts difficult to determine robustly.
We therefore assign four quality flags to our spectroscopic data:
\begin{enumerate}
\item Q\,=\,1 denotes a secure redshift where multiple features were
  identified from bright emission\,/\,absorption lines;
\item Q\,=\,2 denotes a redshift but derived from one or two
  bright emission (or strong absorption) lines;
\item Q\,=\,3 is a tentative redshift based on one (or sometimes two
  tentative) emission or absorption lines.  In these cases, we often
  use the photometric redshift as a guide to identify the line.  These
  redshifts are therefore not independent of the photometric redshifts
  and are thus highlighted in the analysis;
\item Q\,=\,4 is assigned to galaxies with no emission lines or
  continuum detected and so no redshift could be determined.
\end{enumerate}
Examples of spectra from which Q\,=\,1, 2 \& 3 redshifts are
determined are shown in Fig.~\ref{fig:spec1}.  Since the ECDFS has
been the focus of extensive spectroscopic campaigns (although focusing
mainly on bright optical/UV-selected galaxies) six of our ALMA SMGs
have published archival spectroscopic redshifts, and we highlight
these in Table~2.  \footnote{Our goal is to provide a quality
  flag that allows users to gauge the likely success, (or interpret)
  follow up observations a source.  For example, a non-detection of
  the $^{12}$CO emission in a Q\,=\,1 source should be interpreted as
  $^{12}$CO faint, whereas a $^{12}$CO non-detection of a Q\,=\,3
  source may be due to the faintness of the $^{12}$CO emission, or due
  to a misidentified/spurious redshift.}

The emission\,/\,absorption lines we are using to derive redshifts
have a range of physical origins within the galaxies.  For example,
nebular emission lines arise from H{\sc ii} regions and so are
expected to trace the systemic redshift, whereas UV ISM lines can
trace outflowing material and so can be offset from the systemic redshift by
several 100\,km\,s$^{-1}$ \cite[e.g.][]{Erb06c,Steidel10}.
Ly$\alpha$ emission, which is often used to derive spectroscopic
redshifts, also suffers resonant scattering.  As such, to derive
redshifts for each galaxy we adopt the following approach:

%
%
\begin{figure*}
\begin{center}{
\centerline{\psfig{figure=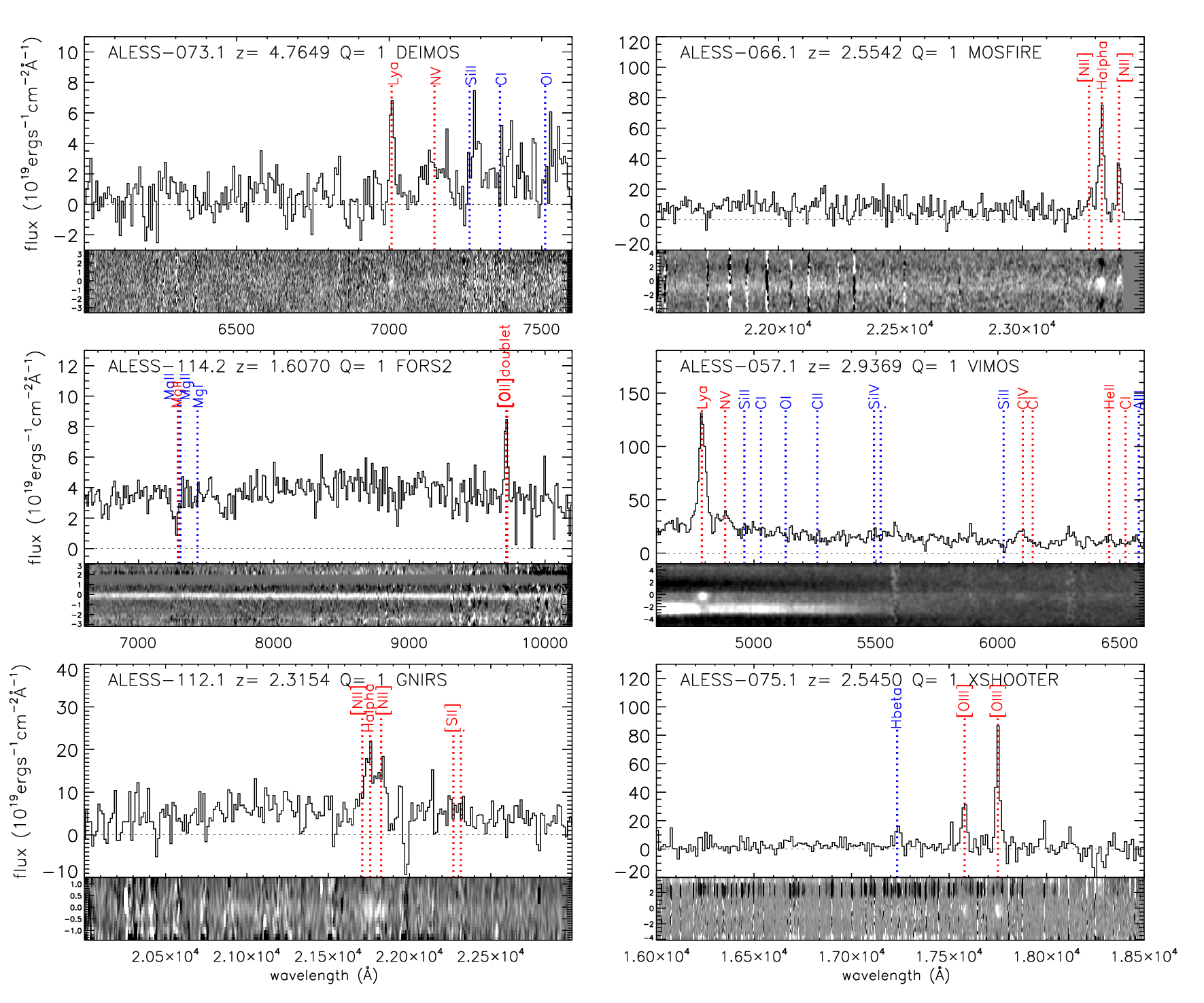,width=7.5in}}
\vspace{-0.5cm}
\centerline{\psfig{figure=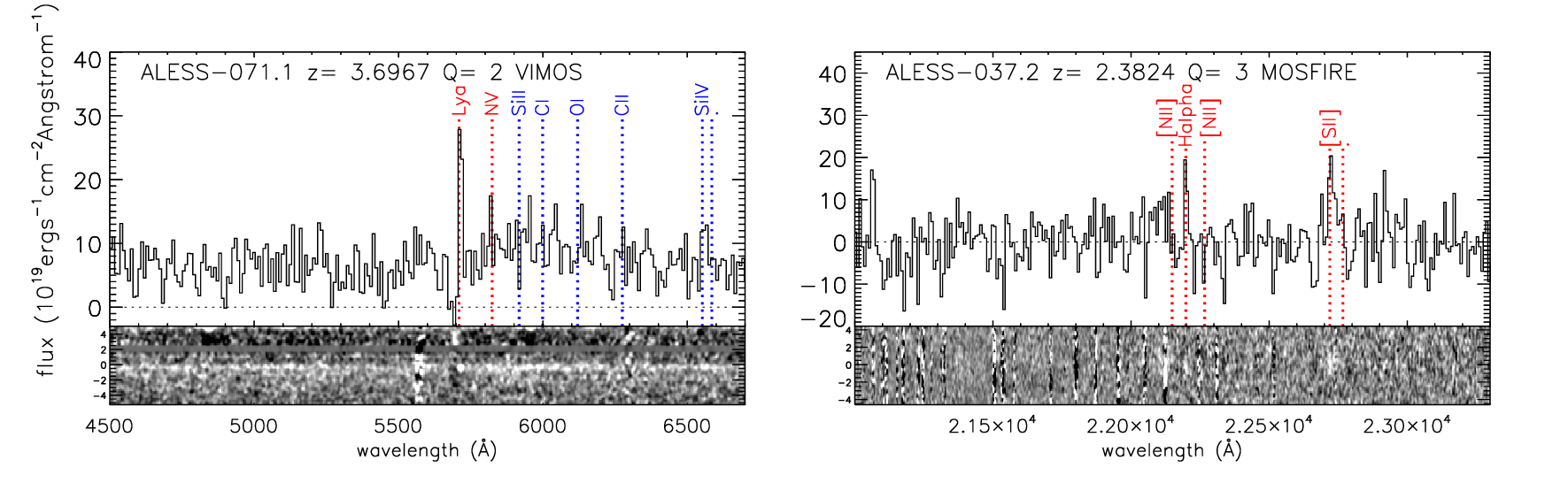,width=7.5in}}
\caption[Q\,=\,1 example spectra of ALESS SMGs]{Example one- and
  two-dimensional spectra of ALESS SMGs from each spectrograph
  used. The upper three rows are high quality (Q\,=\,1) spectra while
  the bottom row shows lower quality examples (Q\,=\,2 and 3 spectra)
  and we mark identified and potential features in all panels, where
  red dashed lines mark typical emission lines and blue dashed lines
  mark typical absorption lines.  In ALESS\,057.1 (an X-ray AGN) the
  bright continuum below the central strong emission line and
  continuum is contamination from higher order emission from an
  adjacent slit on the VIMOS mask. ALESS\,037.2 is an example of a
  Q\,=\,3 redshift where the redshift is determined from narrow
  H$\alpha$, although the apparent ratio of S{\sc ii}\,/\,H$\alpha$ is
  unusually high.}
\label{fig:spec1}
}\end{center}
\end{figure*}

\begin{enumerate}
\item Wherever possible, systemic redshifts are determined using
  nebular emission lines such as H$\alpha$, [O{\sc
      ii}]\,$\lambda\lambda$3726,3729, [O{\sc
      iii}]\,$\lambda\lambda$4959,5007 and/or H$\beta$.  If none of
  these lines are available we use He{\sc ii} or C{\sc
    iii}]\,$\lambda$1909 in emission if they are narrow.
  \item If no nebular emission lines are detected, we determine the
    mean of the redshifts from the UV ISM absorption lines of
    C{\sc ii}\,$\lambda$1334.53, Si{\sc iv}\,$\lambda$1393.76 and
    Si{\sc ii}\,$\lambda$1526.72, or other strong emission lines such as
    N{\sc v}\,$\lambda$1240, Mg{\sc ii}\,$\lambda$2800 and He{\sc ii}.
\item If Ly$\alpha$ is the {\it only} detected line then the redshift
  is determined from a fit to this line, although we caution that the
  velocity offset from the systemic can be up to
  $\sim$\,1000\,km\,s$^{-1}$.  In most of the galaxies where a
  redshift is determined solely from Ly$\alpha$, the observations were
  taken with VIMOS using the low-resolution ($R\sim$\,180) grating,
  precluding any detailed analysis to determine the shape of the
  emission line and judge the influence of absorption on its
observed profile.  Similarly, where possible we avoid using C{\sc
    iv}\,$\lambda$1549 for measuring the redshifts, since it can be
  strongly influenced by winds and frequently exhibits a profile which
  is a superposition of P-Cygni emission and absorption, nebular
  emission and interstellar absorption (or AGN activity).
\end{enumerate}

For the ALESS SMGs, $\sim$\,30\% of the redshifts are determined from a
single line and generally these redshifts are allocated Q\,=\,3 unless
strong continuum features (such as breaks across Ly$\alpha$) are also
identified, which leads to an unambiguous identification and a higher
quality flag.  Single line redshifts are typically backed up by either
continuum breaks across Ly$\alpha$, the absence of other emission
lines that would correspond to a different redshift, line profiles
(i.e.\ asymmetric Ly$\alpha$ profile or identifying the doublet of
[O{\sc ii}]\,$\lambda$3726,3729\AA\ emission).  In seven cases, single
line redshifts are based on detections of Ly$\alpha$; in three cases
they are determined from H$\alpha$ detections in near-infrared spectra
and in five cases they are from detections of the [O{\sc ii}] doublet.

We summarise the main spectroscopic features that we detect in
Table~\ref{tab:specdet} and provide detailed information on each of
the 109 observed SMGs in Table~2.

%
%
\begin{table}
  \small
  \begin{center}{
      \caption{Summary of spectroscopic features}
      \begin{tabular}{ll}
        \hline\hline
        Condition & Number of galaxies \\
        & Total [{\sc supp}] \\
        \hline\hline
        Total & 131 [32] \\
        Q\,=\,1  & 20 [1]  \\
        Q\,=\,2  &  11 [3] \\
        Q\,=\,3  &  21 [3]  \\
        Redshifts measured & 52 [7] \\
        Not observed  & 22 [10] \\
        Observed but no spec $z$  & 57 [15] \\
        & \\
        Ly$\alpha$  &  23 [1] \\[0mm]
        [O{\sc ii}] & 10 [3] \\[0mm]
        [O{\sc iii}] & 6 [0] \\[0mm]
        H$\alpha$  & 14 [3] \\[0mm]
        [O{\sc iii}] \& H$\alpha$ & 3 [0] \\[0mm]
        H$\beta$ & 3 [0] \\
        \hline
      \end{tabular}
      \label{tab:specdet}
    }
  \end{center} 
      {\sc Notes}: The numbers in brackets represents the number of {\sc
        supp} SMGs included in the total in each row.
\end{table}

%
%
\begin{figure}
  \psfig{file=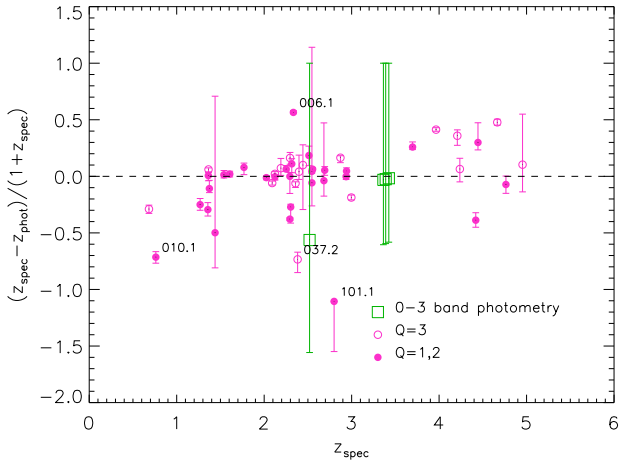,width=3.5in}
  \caption{A comparison of our  spectroscopic redshifts
for ALESS SMGs with their estimated 
    photometric redshifts from \citet{Simpson14}. 
Overall, the photometric redshifts
agree well with our  spectroscopic redshifts with a median
    $\Delta z/(1+z_{\rm spec})$\,=\,0.00\,$\pm$\,0.02.
The errors represent the uncertainties on the
    photometric redshifts determined from the SED fitting in
    \citet{Simpson14}.
We identify those SMGs with  detections
    in just 0--3 photometric bands where the redshift has been determined
    by assuming these SMGs have an absolute $H$-band magnitude
    distribution comparable to that of a complete sample of
    $z\sim$\,1--2 SMGs.  For SMGs with photometric redshift
estimated from only 0--1, and 2--3
    band photometry  are placed at the median
    for those sources of $z\sim$\,4.5 and $z\sim$\,3.5
    respectively.    }
  \label{fig:specz-photz}
\end{figure}

In Fig.~\ref{fig:specz-photz} we compare  our precise spectroscopic 
measurements  for the ALESS SMGs 
to the photometric redshift estimates for these SMGs from
\citet{Simpson14} who determine photometric redshifts for 77 of the
ALESS SMGs which have 4--19 band photometry.  We flag those sources
with spectroscopic redshifts, but poor photometric coverage
and we also highlight the spectroscopic
Q\,=\,3 redshifts  since their spectroscopic
identification is often guided by the photometric redshifts.
Nevertheless, even if these Q\,=\,3 SMGs are omitted, there is good
agreement between the photometric and spectroscopic redshifts with a
median $\Delta z/(1+z_{\rm spec})$\,=\,0.00\,$\pm$\,0.02 and a
variance of $\sigma^2$\,=\,0.1.  In four cases, there appear to be
significant outliers, with $|\Delta z/(1+z_{\rm spec}))|>$\,0.5.  In
these cases, the large offset between the photometric and
spectroscopic redshifts appears to be associated with complex systems
or incomplete photometric coverage, and we briefly discuss these here:

\begin{enumerate}
\item ALESS\,006.1: the photometry of this ALESS SMG appears to
  contaminated by an adjacent low-redshift (and unassociated) AGN, and
  in this case it appears that the SMG is lensed.  The photometry (and
  photometric redshift) is dominated by the foreground AGN.
\item ALESS\,010.1: the Q\,=\,1 spectroscopic redshift is
  significantly lower than predicted by the photometry. There is a
  blue source slightly offset ($<$\,1$''$) from the ALMA position and an
  IRAC source coincident with the ALMA position. {\it HST} imaging
  \citep{Chen15} reveals two galaxies and it is possible that the blue
  source is a lens, as confirmed by high-resolution, $\sim$\,0.1$''$
  ALMA band\,7 follow-up observations; \citep{hodge16}.
\item ALESS\,037.2: the Q\,=\,3 spectroscopic redshift is
  significantly lower than the $z>$\,4 predicted by the
  photometry. However, the spectroscopic redshift is based on two
  tentative line detections at the correct separation for H$\alpha$
  and [S{\sc ii}] (see Fig.~\ref{fig:spec1}; [N{\sc ii}], if present
  would lie under strong sky lines) and the photometric redshift is poorly
  constrained and based on detections in six bands and limits in a
  further six. Furthermore, the spectroscopic line identifications
  would not correspond to any common emission lines if the photometric
  redshift is correct.
\item ALESS\,101.1: this has a Q\,=\,2 redshift based on a single
  detection of Ly$\alpha$. It has poor constraints on the photometric
  redshift with photometric detections in only five bands and no
  detections below $J$-band. Thus the spectroscopic redshift is
  significantly more reliable.
\end{enumerate}

For a significant fraction of the ALMA sample targeted
in our survey, we were unable to
derive a spectroscopic redshift (these are assigned $Q$\,=\,4 in
Table~2).  To understand  whether this is caused by magnitude
limits or their redshifts,  we first compare the photometric redshifts
of the spectroscopic failures to those for the SMGs for which we were
able to determine a spectroscopic redshift.  The median photometric
redshift of spectroscopic failures is $z=$\,2.4\,$\pm$\,0.2,
compared to $z=$\,2.4\,$\pm$\,0.1 for the sources for which we were
able to measure a spectroscopic redshift
 (these estimates use the best-fit photometric
redshifts values, but they change by less
than the quoted uncertainty if the full photometric redshift probability distributions
are used instead). 
This suggests that the SMG with
spectroscopic failures are not  at {\it much} higher
redshifts than those SMGs where we have succeeded in obtaining a
redshift. 
Similarly there does not appear to be any correlation with submillimetre
flux:  for the 52 SMGs with spectroscopic
redshifts, the median 870-$\mu$m flux is $S_{870\mu \rm
  m}$\,=\,$4.2^{+0.3}_{-0.4}$\,mJy, whereas  those 57 SMGs where we could
not determine a redshift have a median $S_{870\mu \rm
  m}$\,=\,$4.3^{+0.2}_{-0.6}$\,mJy.

%
%
\begin{figure*}
  \begin{center}
    \psfig{figure=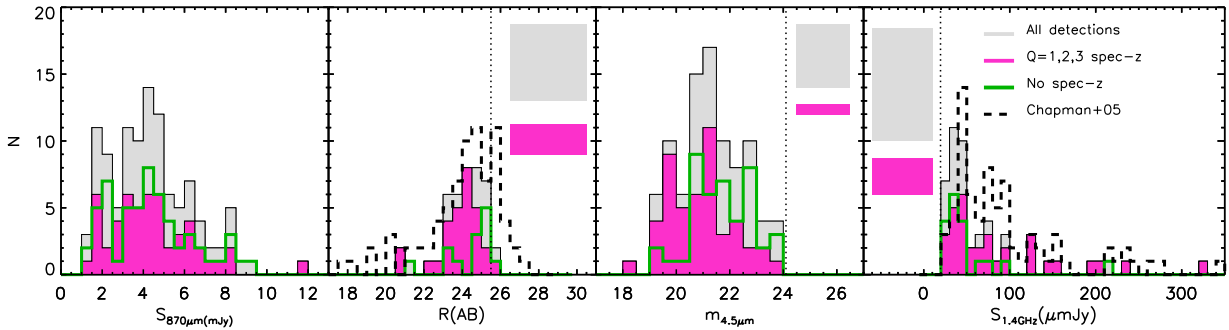,width=7.in}
    \caption{Fundamental observable properties of our 
spectroscopic sample of SMGs, comprising 870\,$\mu$m fluxes,
$R$-band and 4.5\,$mu$m magnitudes and 1.4\,GHz fluxes.  The
distributions are 
compared to those of the parent population of ALESS SMGs
(where the parent sample comprises the 109/131 SMGs
      that were targeted in our spectroscopic survey).  In
all panels we show three
distributions:  for 
the full sample (with and without spectroscopic redshifts);
 the properties of the SMGs with Q\,=\,1, 2 or 3
      spectroscopic redshifts and the distribution for SMGs with
      photometry but no spectroscopic redshift. 
As separate boxes we also indicate the proportion of the
full and spectroscopic samples which are below the detection limit
of the observations in each waveband (these 3$\sigma$ detection
      limits are indicated by dotted lines in each panel). 
On average, we find that the SMGs for which
      we were able to determine a redshift are marginally brighter in
      the $R$-band, and m$_{\rm 4.5\mu m}$ than those for which we were
      unable to determine a redshift, however, the likelihood of
      determining a redshift is independent of the 870\,$\mu$m
      flux density and so our survey is unbiased in this regard.
In addition in the $R$-band and 1.4\,GHz panels we
also show the equivalent distribution for
the spectroscopic sample of 73 radio-identified submillimetre
sources from Chapman et al.\ (2005), which exhibit comparable
properties to our sample.
Note that ALESS\,020.1 has a
      very bright radio flux  of $\sim$\,4.2\,mJy and is therefore
      not shown on the 1.4\,GHz panel.  }
    \label{fig:mags}
  \end{center}
\end{figure*}

Next, we test the hypothesis that we were unable to measure
spectroscopic redshifts for some ALMA SMGs simply due to their faint optical magnitudes.  In
Fig.~\ref{fig:mags} we show the distributions of the $S_{870\mu \rm
  m}$ flux density, $R$-band and 4.5$\mu$m magnitudes and
1.4\,GHz flux density for the 109 (out of 131) ALESS SMGs that were
spectroscopically targeted.  The median $R$-band magnitude of the
ALESS SMGs with spectroscopic redshifts is $R=$\,24.0\,$\pm$\,0.2
whereas the median magnitude of those SMGs for which we could not
measure a redshift is $\sim$\,1 magnitude fainter, at
$R=$\,25.0\,$\pm$\,0.4.  Turning to longer wavelengths, in the
mid-infrared, the median magnitude at 4.5\,$\mu$m is $m_{4.5\mu \rm
  m}=$\,20.9\,$\pm$\,0.2 for the ALESS SMGs with spectroscopic
redshifts, as compared to a median of $m_{4.5\mu \rm
  m}=$\,21.7\,$\pm$\,0.2 for those targeted SMGs for which we could
not derive a spectroscopic redshift.
Hence, there is evidence that the ALESS SMGs for which we were unable to determine a
spectroscopic redshift are marginally fainter in $R$ and $m_{4.5\mu
  \rm m}$ than those for which we were able to measure a
spectroscopic redshift 
(and also may have slightly redder $R-m_{4.5}$ colours).

\begin{figure}
  \psfig{file=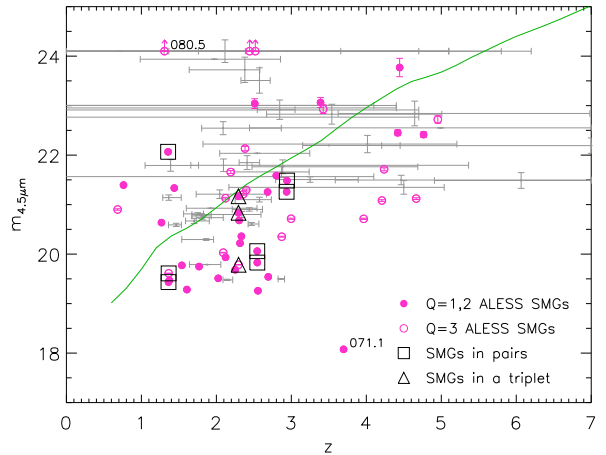,width=3.5in}
  \caption{A plot showing the distribution of  
 4.5\,$\mu$m apparent magnitude  versus redshift for ALESS SMGs. 
We see a tendency for more distant SMGs to have fainter 4.5\,$\mu$m
magnitudes and to assess this we plot a line showing the expected variation with
    redshift for a galaxy with a fixed, non-evolving  luminosity, assuming the
    composite ALESS SED from \citet{Simpson14} (see also
    Fig.~\ref{fig:sed}).  This track is normalised to the median
    apparent magnitude in 4.5\,$\mu$m at a median redshift of
    $z=$\,2.4.  The data roughly follow this trend, although they exhibit
at least an order of magnitude variation in 4.5\,$\mu$m magnitude
at a fixed redshift.
Those SMGs which are found to be physically
    associated (pairs or triples) with other SMGs are highlighted.
    Those in associations have a marginal tendency to be among the brighter SMGs (and
    therefore could potentially be more massive; see
    \S~\ref{sec:smg_env}).  Photometric
    redshifts (where spectroscopic redshifts are not available) are
    shown as their $\pm 1\sigma$ ranges given in \citet{Simpson14} and
    Table~2. The two extreme outliers are identified with their ALESS
    ID.}
  \label{fig:hubble}
\end{figure}

In Fig.~\ref{fig:hubble} we plot the
redshifts of the ALESS SMGs versus their 4.5\,$\mu$m apparent
magnitudes.  At the typical redshift of SMGs ($z\sim$\,2.4), the
4.5\,$\mu$m emission provides the most reliable tracer of the underlying
stellar mass, since it corresponds to rest-frame $\sim$\,1.6\,$\mu$m ($H$-band).
As a guide, to crudely test how the 4.5\,$\mu$m magnitude depend on
redshift in our sample, we generate a non-evolving starburst track,
based on the composite SED for the ALESS SMGs (shown in
\citealt{Simpson14} but updated to contain the spectroscopic redshift
information in Fig.~\ref{fig:sed}).  
This model SED has been normalised  to
the median apparent 4.5\,$\mu$m magnitude for the spectroscopic and photometric redshift samples  at
the median redshift of $z\sim$\,2.4. 
The dependence of 4.5\,$\mu$m flux
with redshift for our spectroscopic sample is consistent with this
track, although with a spread of $\sim$\,2 magnitudes at fixed
redshift.  However, the data do show a trend of decreasing 4.5\,$\mu$m
flux with increasing redshift.  \citet{Smail04} \citep[see
  also][]{Serjeant03} also identified a similarly large spread in
$K$-band magnitudes for SMGs. 

Hence we see both a spread in the apparent rest-frame near-infrared luminosities within the SMG
population, as well as the  fainter optical apparent magnitudes (and
redder colours) for those
SMGs which we failed to obtain redshifts for and  marginally higher
photometric redshifts compared to those for which spectroscopic redshifts
were measured.  
Each of these trends are  weak, but they do
suggest several factors may be driving the spectroscopic
incompleteness:  a range in stellar masses for SMGs at
a fixed redshift (a demonstration of the diversity of the SMG
population),  varying levels of strong dust extinction and fainter
apparent optical fluxes for SMGs at higher redshifts
(due to the K correction and 
increasing distance).

In terms of the radio-detected sub-sample, from the entire {\sc
  main+supp} ALESS catalogue, 53\,/\,131 ALESS SMGs are radio-detected,
and we have targeted 52 with spectroscopy, measuring redshifts for 34.
The median 1.4\,GHz flux density of the SMGs with spectroscopic
redshifts is $S_{1.4\rm GHz}$\,=\,63\,$^{+12}_{-13}\mu$Jy compared to
$S_{1.4\rm GHz}$\,=\,39\,$^{+6}_{-2}\mu$Jy for those without
spectroscopic redshifts (Fig.~\ref{fig:mags}).  Thus, SMGs for which
we were unable to determine a spectroscopic redshift are fainter at
radio wavelengths than those for which we measured a spectroscopic
redshift.

\section{Spectroscopic redshift distribution}
\label{sec:specdist}

\begin{figure*}
\begin{center}{
\psfig{figure=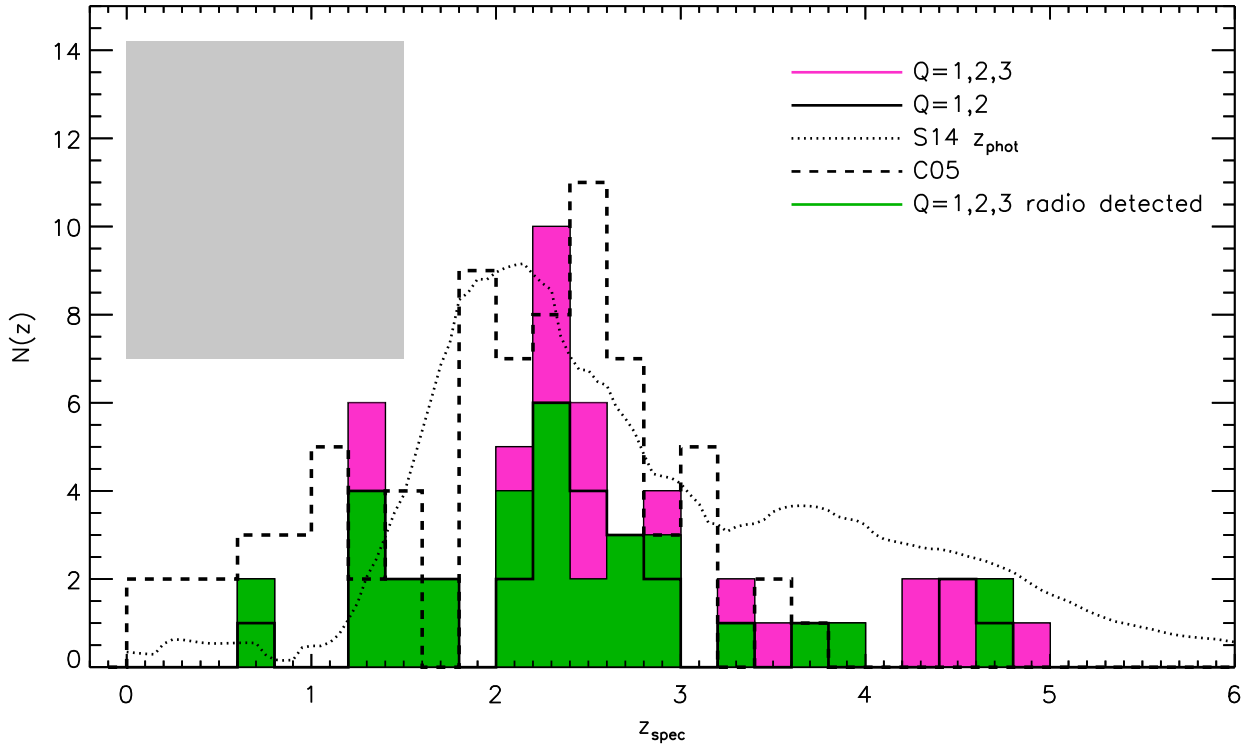,width=7.in}
\caption{The spectroscopic redshift distribution of the SMGs from our
survey.  Those SMGs with secure redshifts
  (Q\,=\,1 \& 2) are shown, as well as the distribution for
all Q\,=\,1, 2 \& 3 redshifts.  We
  compare the distribution to the probability density function of the
  photometric redshifts from \citet{Simpson14} normalised to the same
total number of sources.  We also compare to
  the  redshift distribution of radio-identified submillimetre
sources from \citet[C05][]{Chapman05}.  We see very striking
differences between the ALESS SMG redshift distribution and 
that for \citet{Chapman05}, both at low and high redshifts, $z\lsim$\,1 and
$z\gsim$\,3.5. In particular
  the ALESS SMGs have a  spectroscopic redshift distribution that extends to higher
  redshift, with $\sim$\,23\% of the SMGs at $z>$\,3 and an even larger
proportion in the more complete, but less precise, photometric redshift distribution from \citet{Simpson14}.  
 To mimic the selection of the radio-identified \citet{Chapman05} sample, 
the redshift distribition of the radio-detected ALESS SMGs
  are highlighted.  This shows that there are still discernable differences
  between the redshift distributions of the radio-detected ALESS SMGs
  and those from  \citet{Chapman05}  at low redshifts, $z\lsim$\,1,
raising the possibility that some of the low-redshift radio counterparts
to submillimetre sources claimed by \citet{Chapman05} could be misidentifications.  
The bin size is $\Delta z=$\,0.2 and the  grey shaded box indicates the incompleteness in the
Q\,=\,1, 2 \& 3 sample compared to the parent population of targeted SMGs in the field.  
}
\label{fig:Nz}
}\end{center}
\end{figure*}

The spectroscopic redshift distribution of the ALESS SMGs is shown in
Fig.~\ref{fig:Nz}.  In total 52 redshifts have been determined for the
ALESS SMGs: 45 {\sc main} catalogue SMGs and seven {\sc supp} catalog
SMGs.  We also overlay the probability density function of the
photometric redshift distribution of ALESS SMGs from
\citet{Simpson14}, scaled to the same number of sources.  The Q\,=\,1
\& 2 and Q\,=\,1, 2 \& 3 distributions are shown as individual
histograms to test the effect of including the Q\,=\,3 redshifts. The
full redshift distribution ranges between $z$\,=\,0.7--5.0, with a
significant (but not dominant) tail at $z\geq $\,3 for those
distributions without a radio-selection.

\begin{figure}
  \psfig{figure=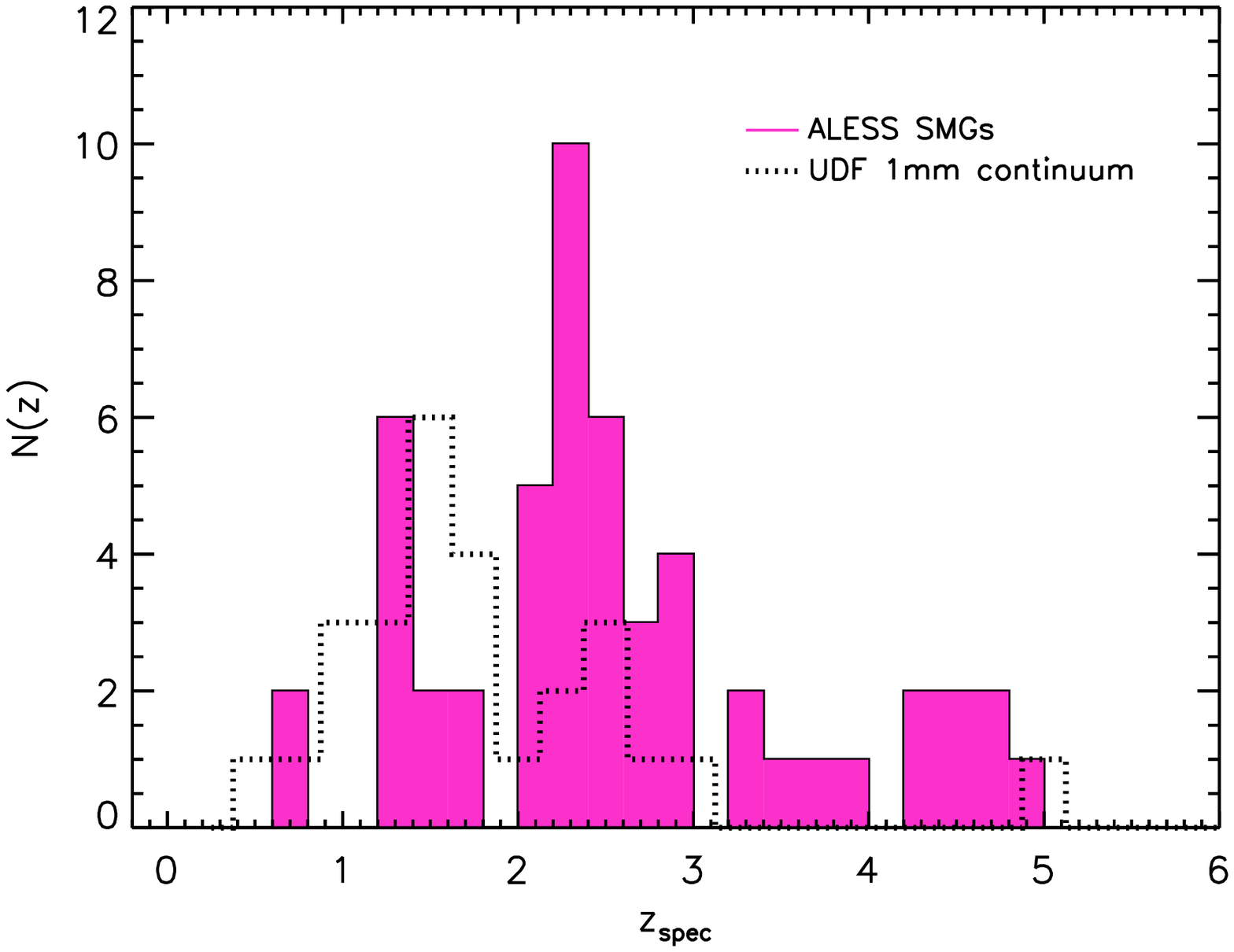,width=3.5in}
\caption{The spectroscopic redshift distribution of the SMGs in our
  870\,$\mu$m survey compared to that for two faint 1.1-mm selected
  samples in the UDF from \citet{Aravena16} and \citet{Dunlop16} (we
  note that the total number of sources for the distributions shown
  are not the same).  These SMG samples have quite different selection
  functions and levels of incompleteness and so we do not draw any
  strong conclusions from the apparent differences between them,
  beyond noting that both distributions peak at relatively high
  redshifts, $z\sim$\,1.5--2.5, and reach out to $z\sim$\,5 with the
  more numerous ALESS 870\,$\mu$m sample showing a more significant
  high-redshift tail beyond $z\sim$\,3.  }
\label{fig:Nzcomp}
\end{figure}

In Fig.~\ref{fig:Nzcomp} we show the ALESS spectroscopic redshift
distribution and compare this with the 1.1-mm selected (U)LIRGs from
the recent ALMA surveys of the {\it Hubble} Ultra Deep Field (UDF) by
ASPECS \citep{Aravena16,Walter16} and \citet{Dunlop16} Given the
different selection wavelengths, flux limits and sample sizes between
the ALESS SMGs and the ALMA\,/\,UDF galaxies, we caution against
drawing strong conclusions about the differences between these
redshift distributions \citep[for a detailed discussion
  see][]{Bethermin15}.  Nevertheless, we note that all of these
distributions peak at $z\sim$\,2.0\,$\pm$\,0.5, with a suggestion that
fainter sources may lie at lower redshifts on average.

Before continuing with the analysis, we briefly assess the effect on
our sample of including the {\sc supp} SMGs and those with only
Q\,=\,3 redshifts.  \citet{Karim13} demonstrate that up to $\sim$\,30\%
of the {\sc supp} sources are likely to be spurious.  However, 
{\sc supp} sources which have an optical\,/\,near-infrared counterpart
have a lower liklihood of being spurious sources.  The median redshift
of the {\sc main} catalogue SMGs with Q\,=\,1, 2 \& 3 redshifts is
$z=$\,2.5\,$\pm$\,0.1 with an interquartile range of
$z=$\,2.1--3.4, whereas the median redshift of the {\sc main}+{\sc
  supp} catalogue with Q\,=\,1, 2 \& 3 redshifts is
$z=$\,2.4\,$\pm$\,0.1 with an interquartile range of
$z=$\,2.1--3.0.  The median redshift of the Q\,=\,1, 2 \& 3 SMGs in
the {\sc supp} sample alone is $z=$\,2.3\,$\pm$\,0.5.  Thus, the
median redshifts of these various samples are all consistent.  Indeed,
a two-sided Kolmogorov-Smirnov (K-S) test between the {\sc main} and
{\sc supp} samples suggests only a 60\% likelihood that they are
drawn from different populations.  Since the statistics of the samples
do not vary strongly with the inclusion of the {\sc supp} sources, we
are therefore confident that including the {\sc supp} sources in our
analyses is unlikely to  bias any of our results.

Since most previous SMG redshift surveys have, by necessity, relied on
radio detections to identify probablistically the likely counterparts, we briefly
discuss the properties of the radio-detected subset of the ALESS SMGs,
as this provides a reasonable comparison to previous work. 
In our sample we targeted 52 of the 53 radio-detected SMGs with
spectroscopy and measured redshifts for 34 of them (65\%).  The median
1.4\,GHz radio flux density of the 34 radio-detected ALESS SMGs with
spectroscopic redshifts is 63\,$^{+12}_{-13}\mu$Jy, as compared to
50\,$^{+6}_{-5}\mu$Jy for {\it all} 52 radio-detected SMGs.
In contrast, the median radio flux density of the 73 radio-detected
submillimetre sources in \citet{Chapman05} with spectroscopic redshifts is
75$^{+8}_{-3}\mu$Jy.  On average, the radio-detected ALESS SMGs with
redshifts are $\sim$\,20\% fainter at 1.4\,GHz than the
\citet{Chapman05} sample and our spectroscopic completeness is $\sim$\,10\%
lower.
We note that it appears that the \citet{Chapman05} radio-identified
submillimetre sources have a higher
AGN fraction than our ALESS sample, and indeed up to $\sim$\,40\% of
their  sample exhibit signatures of AGN activity in the X-rays,
spectra or from their broad-band optical/mid-infrared SEDs
\citep[e.g.][]{Alexander08,Hainline11}.  \citet{Wang13}
find an AGN fraction of $\sim$\,17$^{+16}_{-6}$\% for the ALESS SMGs.
Typically AGN spectra have stronger, more easily identifiable emission
features and thus our $\sim$\,10\% lower spectroscopic completeness may be due
to a lower AGN fraction.

%
%

%
%

\section{Discussion}
\label{sec:discuss2}

Although the primary aim of this work is to determine the redshifts of
unambiguously identified SMGs to support further detailed follow-up
(e.g.\ CO or H$\alpha$ dynamics, \citealt[e.g.][]{Huynh13}), there is also a wealth of
information contained within the spectra themselves concerning the
dynamics, chemical composition, and energetics of these SMGs.
Furthermore, the redshifts can be used as constraints in SED models
(e.g.\ constraining the star-formation history and so the stellar
masses) and to investigate the environments in which these SMG reside.

\subsection{Spectral diagnostics}

\subsubsection{Stacked spectral properties}
\label{sec:stack}
\begin{figure*}
  \begin{center}
      \psfig{figure=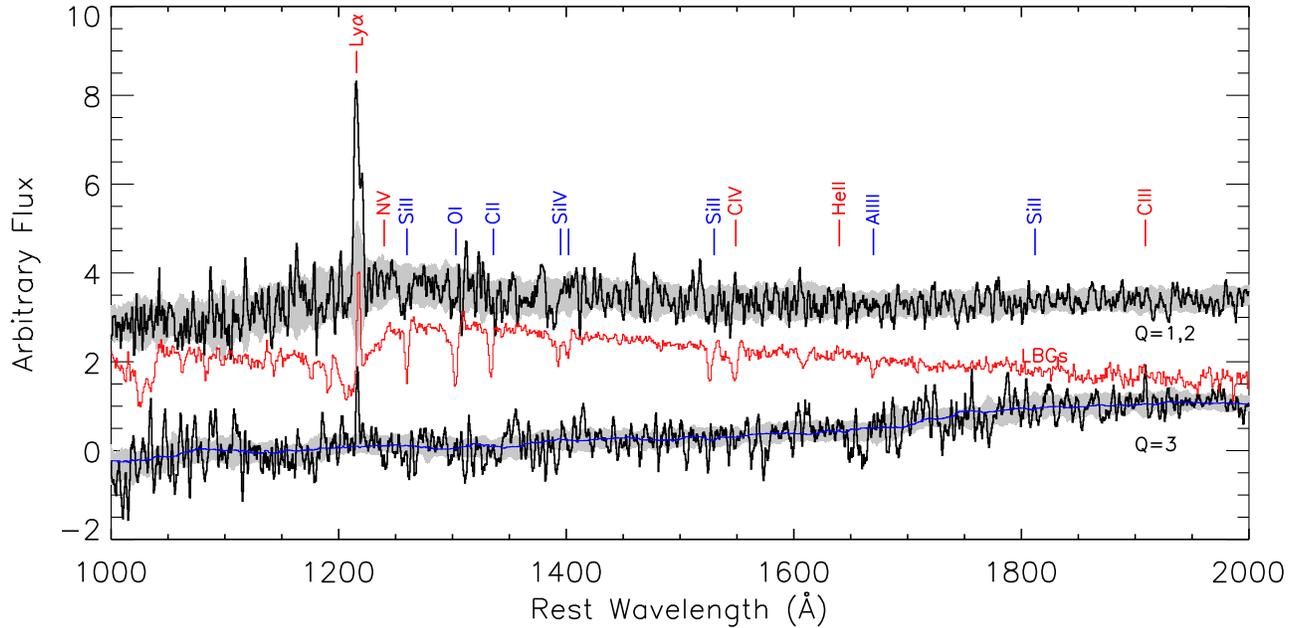,width=7.in}
      \caption{Composite spectra around the Ly$\alpha$ emission line
        ($\sim$\,1215{\AA}). The spectra are averaged and weighted by
        their noise and the uncertainty is derived from bootstrap
        resampling of the spectra included in the stack and is shown
        as the grey shaded regions. The upper spectrum shows the stack
        of all the Q\,=\,1 and 2 spectra which shows a number of
        potential absorption features, as well as the Ly$\alpha$
        emission line.  For comparison the composite spectrum of LBGs
        from \citet{Shapley03} is overlaid in red (and offset for
        clarity).  The Q\,=\,3 stack at the bottom was produced to
        test the validity of the uncertain Q\,=\,3 redshifts by
        identifying features in their composite spectrum. The solid
        blue line is a running median of the Q\,=\,3 composite. We see
        apparently significant detections of Ly$\alpha$ and a weak
        feature which may be C{\sc iii}]\,$\lambda$1909 in the Q\,=\,3
      composite, which if real may indicate that some of these
      redshifts are correct. }
      \label{fig:lya}
  \end{center}
\end{figure*}

\begin{figure*}
\psfig{file=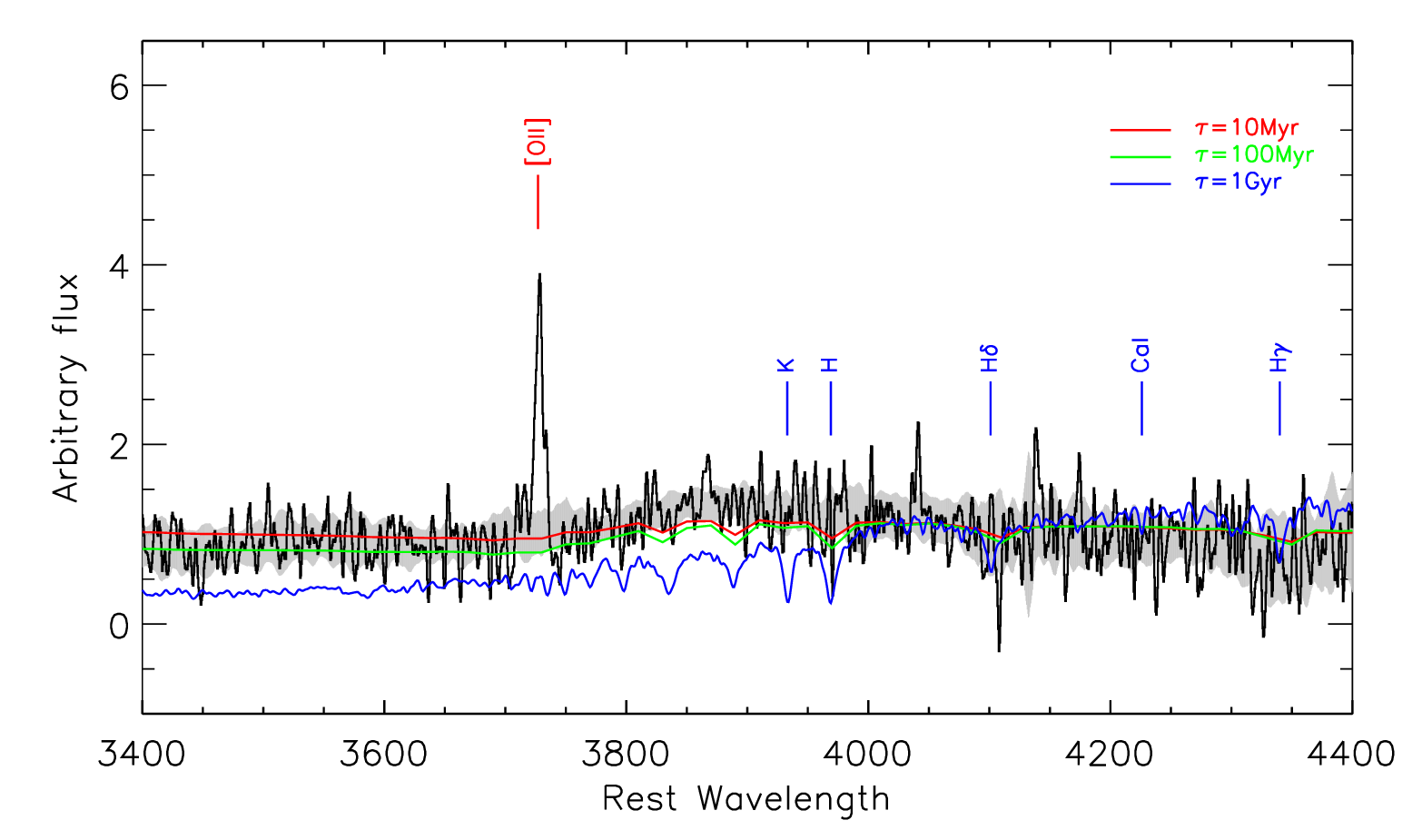,width=3.7in}
\psfig{file=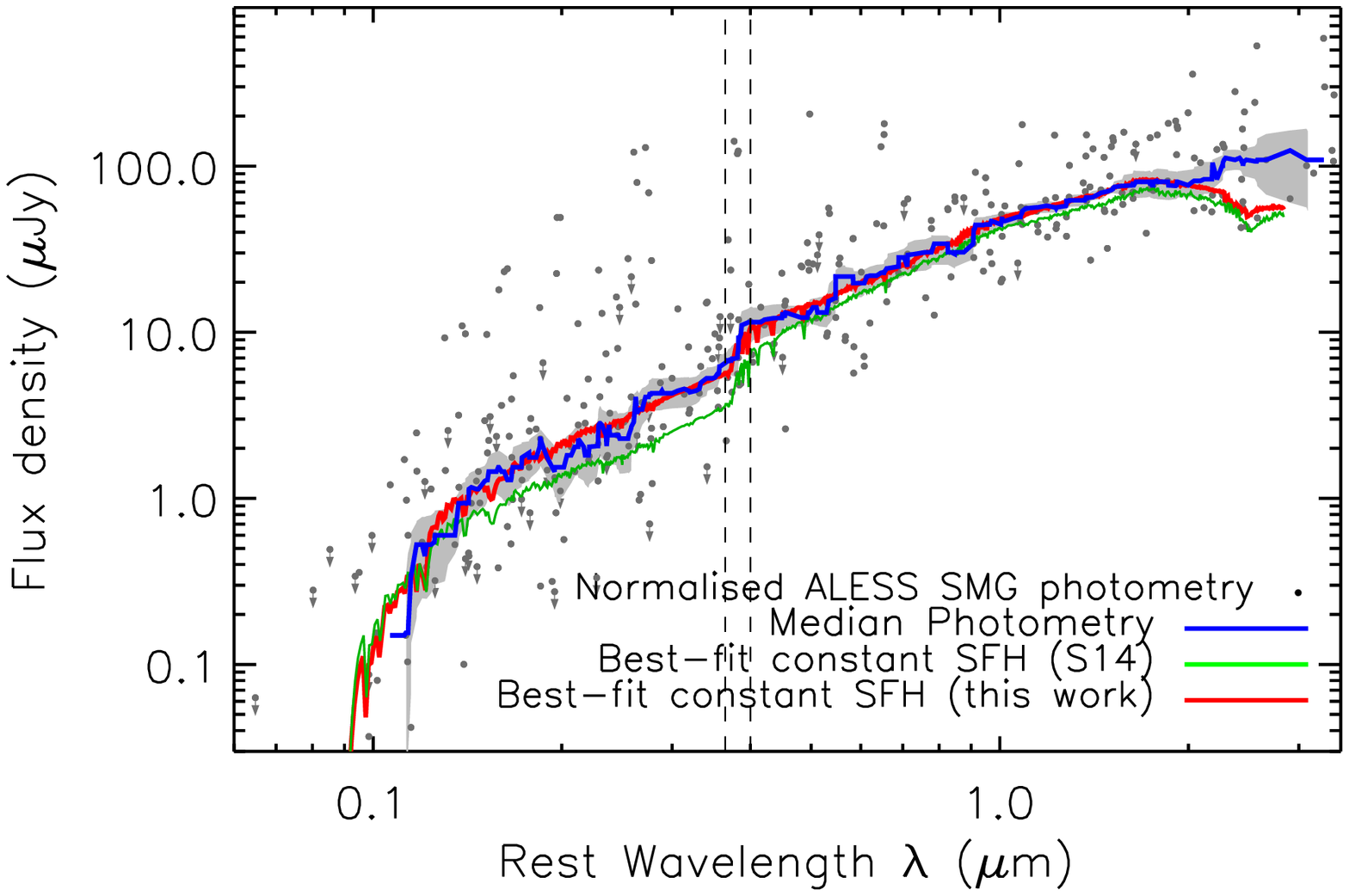,width=2.1in,angle=90}
\caption{{\it Left:} The composite spectrum covering restframe
  3400--4400{\AA} of the Q\,=\,1 \& 2 ALESS spectra with the X-ray AGN
  removed from the sample.  This shows strong [O{\sc ii}] emission and
  potentially H$\delta$ absorption, as well as the presence of a
  spectral break around $\sim$\,3800\AA resulting from the Balmer
  series.  We overlay model spectra for a continuous 100\,Myr
  starburst observed at 10, 100 and 1000\,Myr.  The model spectra for
  the 10\,Myr burst provides the closest match to the strength of the
  Balmer break.  The spectra were normalised by their median continuum
  flux between 2900--3600{\AA} and sky-subtracted by the same method
  as in Fig.~\ref{fig:lya}. We show the uncertainty in the composite
  derived from a bootstrap resampling of the sources included in the
  composite as a grey shaded region. {\it Right:} A composite SED
  using the photometry from Simpson et al.\ (S14, 2014) for those
  ALESS SMGs with Q\,=\,1, 2 \& 3 spectroscopic redshifts. The
  photometry for each sources has been de-redshifted and normalised by
  their rest-frame $H$-band luminosity. The solid line represents the
  running median of 20 points per bin. The shaded region indicates the
  bootstrap error on the running median.  The red curve represents the
  best fit model SED assuming a constant star-formation rate to the
  average photometry for all ALESS SMGs, whereas the green curve is
  the equivalent model fit taken from Simpson et al.\ (2014). The
  de-redshifted photometry and limits are shown as grey points and
  arrows respectively. The vertical dashed lines indicate the Balmer
  (3646\AA) and 4000{\AA} breaks.}
\label{fig:sed}
\end{figure*}

Stacked spectra are a useful tool to detect weak features that
are not visible in individual spectra and also for determining the average
properties of the population.  We therefore produce composite
spectra over two different wavelength ranges, one covering
Ly$\alpha$ and UV ISM lines and one around the [O{\sc ii}]$\lambda$3727 and
Balmer break, and we use these
to search for evidence of  emission/absorption features and
continuum breaks.  To construct the composites, we first transform
each spectrum to the rest-frame using the {\it best} redshift in
Table~2.  Where the sky subtraction leaves significant
residuals, the region within $\pm$\,5{\AA} of the sky lines are masked
before stacking (and we use the OH line catalogue 
from \citet{Rousselot00} to identify the bright sky lines in the
near-infrared).
We then sum the spectra, inverse weighted by the noise (measured as
the standard deviation in the region of continuum over which they have
been normalised).  In the case of the 1000--2000{\AA} composite
(Fig.~\ref{fig:lya}), we normalise the spectra by their median
continuum value at $>$\,1250{\AA} and in the case of the composite
around 3400--4400{\AA} (Fig.~\ref{fig:sed}), we normalise by the
median continuum value between 2900--3600{\AA}.  We note that when
transforming the spectra to the rest-frame, in a number of cases, the
UV ISM lines and Ly$\alpha$ can be significantly offset in velocity
from this systemic redshift (see Fig.~\ref{fig:zoffsets}).  In the
composite spectrum these spectral features may therefore appear broadened
and offset.

We first discuss the composite spectra of the region around Ly$\alpha$,
1000--2000\AA, see 
Fig.~\ref{fig:lya}.   We show a composite
constructed from just the Q\,=\,1 and 2 spectra
which displays strong Ly$\alpha$ and a
continuum break at $\sim$\,1200{\AA}.  The spectrum also shows 
two Si{\sc ii} absorption lines and apparently offset Si{\sc iv} absorption,
as well as potentially weak C{\sc iv} absorption and emission and O{\sc i}
absorption.  If the feature identified as Si{\sc iv} is real, then it and 
the weaker C{\sc iv} features, both of which show 
  blueshifts, may be indicative of strong stellar
  winds.  To illustrate
the typical strength of the absorption
features we also overlay the composite spectrum of $\sim$\,200 Lyman
  break galaxies (LBGs) from \cite{Shapley03} (the LBG composite shown
  here corresponds to the quartile of 200 LBGs from the
  \citet{Shapley03} sample that has the closest match in Ly$\alpha$
  equivalent width to our ALESS sample).  
We note that due to the different wavelength ranges of the different
instruments used and the fact that we de-redshift and stack in the
rest-frame, not all the spectra in our stack contribute to the full
wavelength range.

We also construct a composite from the Q\,=\,3 spectra and plot this
in Fig.~\ref{fig:lya}.  The purpose of this is to test the reliability
of the redshifts derived for the Q\,=\,3 spectra by searching for weak
spectral features which are undetected in the individual spectra, but
become visible in the stacked spectrum due to the improved signal to
noise.  In addition to an emission line identified as Ly$\alpha$
(which is frequently the feature used to derive the redshift for these
sources), we see only a potential emission feature which would
correspond to C{\sc iii}]$\lambda$\,1909 and no evidence of a break in
  the continuum across the bluer emission line.  If the C{\sc
    iii}]$\lambda$\,1909 emission is real, then it may indicate that
    some of the Q\,=\,3 redshifts are correct.

%
%

To search for continuum breaks and  absorption lines in the
rest-frame optical, and to 
determine if we can constrain the luminosity weighted age of the
stellar populations in SMGs, we also produce a rest-frame composite of the
Q\,=\,1 and 2 spectra over the wavelength range of 3400--4400{\AA}
(removing the bright X-ray AGN from the sample; \citealt{Wang13}) and
show this in Fig.~\ref{fig:sed}.  We detect strong [O{\sc ii}],
and potentially also H$\delta$ absorption (Fig.~\ref{fig:sed}).  In addition, we
see in this composite that
continuum falls off bluewards of $\sim$\,3800{\AA}.  A break in
this region could be due to the 4000\AA\, break, typically observed in
older stellar systems, or more likely the Balmer break at $\sim$\,3656{\AA}.  The
Balmer break arises in stellar populations which are either
experiencing on-going star formation over the previous $>$\,100\,Myr, or
in post-starburst stellar populations, 0.3--1\,Gyr after the
strongest star formation has ended \citep{Shapley11}.  In the composite, 
the position discontinuity is more consistent with the Balmer
break than a 4000{\AA} break, as the continuum at 3500--3600{\AA}
is $(1.5\pm0.1)\times$ lower than it is at 3900--4000\AA.

To try to place limits on the age of the 
visible stellar populations within the ALESS SMGs, we
use the SED templates from \citet{Bruzual03} to predict the 
spectra expected from a starburst of 100\,Myr duration observed at ages
of 10\,Myr,
100\,Myr and 1\,Gyr (post-starburst).  We redden the model spectra
using the reddening law from \citet{Calzetti00} adopting the median
extinction of A$_{\rm V}$\,=\,2 for the ALESS SMGs, as derived from
SED fitting (see \S~\ref{sec:magphys}).  As Fig.~\ref{fig:sed} shows,
the stellar continuum emission
seen in the composite spectrum is most similar to an on-going burst
(i.e.\ undergoing star-formation on 10--100\,Myr timescales), as
expected for these strongly star-forming galaxies.

As well as stacking the spectra, we can also create a rest-frame
broad-band SED for a ``typical'' SMG (or at least ``typical'' of 
the brighter/bluer examples for which redshifts can be measured).  \citet{Simpson14} and \citet{Swinbank14} discuss the
optical\,/\,near-infrared and far-infrared\,/\,radio photometry of the
ALESS SMGs \citep[see also][]{DaCunha15}.  By combining the
multi-wavelength photometry with spectroscopic redshifts for the 52 ALESS
SMGs, we create composite SEDs from the rest-frame UV to radio
wavelengths.  First, we transform the photometry to the rest-frame,
and then stack the photometry (normalised by rest-frame $H$-band
luminosity; see \S5.1.2).  A running median is then calculated through the data to
produce an average SED which we show in Fig.~\ref{fig:sed}.   We also
overlay a {\sc hyper-z} fit using a constant star-formation history,
which indicates (as expected) a heavily dust reddened spectrum of
these SMGs.  Our best-fit constant star-formation model shows a slightly
bluer continuum than that derived using the photometric redshift sample
by \citet{Simpson14}, illustrating a modest bias to bluer restframe UV
continuua in those SMGs for which we can measure spectroscopic redshifts
for.  Nevertheless, our spectroscopic composite SED still display a
very red continuum shape and a clear break at $\sim$\,3800\AA, as seen
in the composite spectrum at this wavelength (Fig.~\ref{fig:sed}).

%
%

\subsubsection{UV-to-radio SEDs}
\label{sec:magphys}

\begin{figure}
\begin{center}{
\psfig{figure=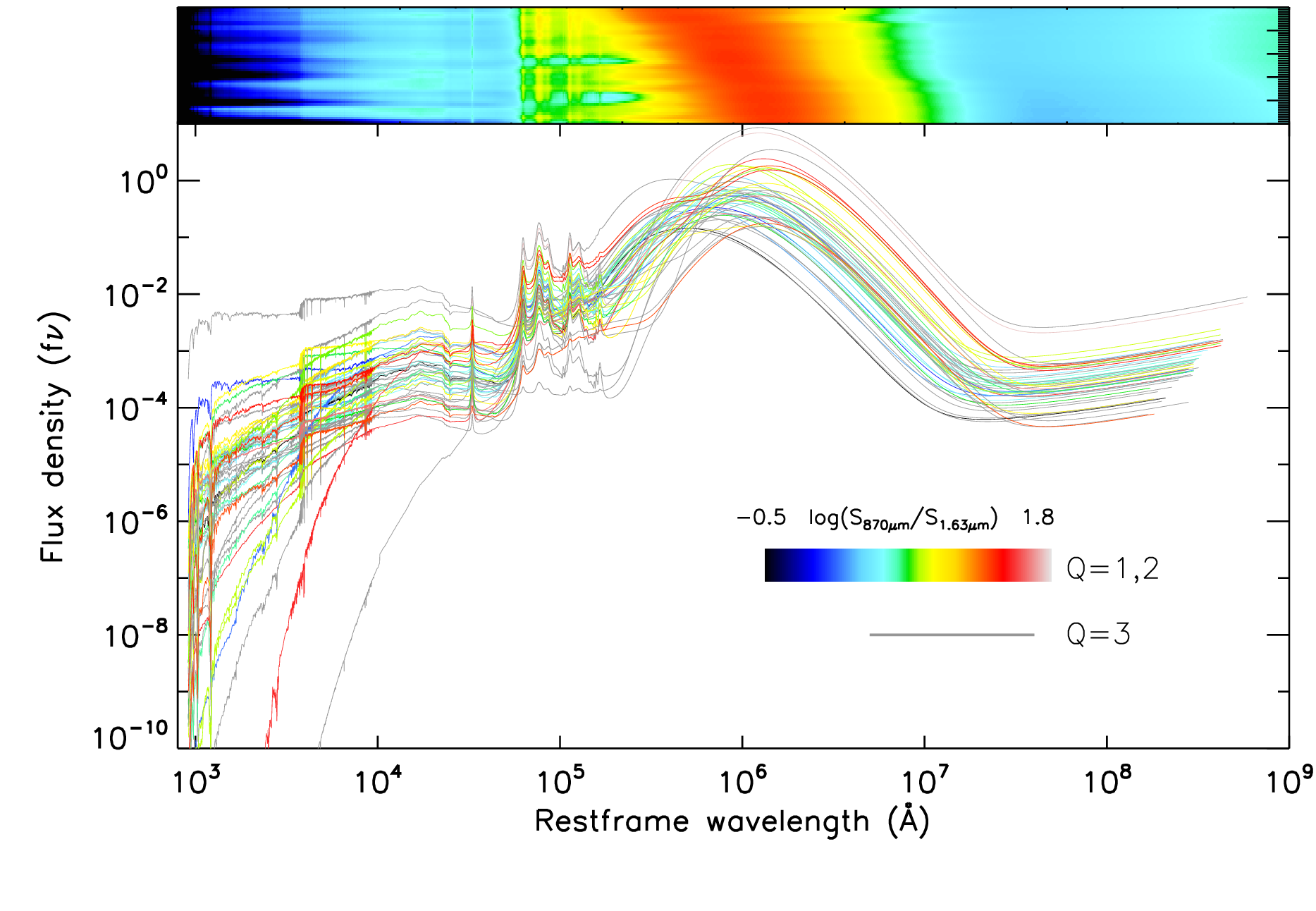,width=3.5in}
\caption{The best-fit rest-frame SEDs for all ALESS SMGs with
  spectroscopic redshifts.  These SEDs have been fitted using {\sc
    magphys} \citep[see][]{DaCunha08} and are normalised by their
  far-infrared (8--1000\,$\mu$m) luminosity.  The coloured curves
  represent SEDs for SMGs with Q\,=\,1 \& 2 redshifts. They are
  colour-coded by the logarithm of their ratio of rest-frame
  S$_{870\mu\rm m}$/$H$ flux density (with red denoting a higher
  ratio).  Grey curves represent SEDs for SMGs with Q\,=\,3 redshifts.
  We see a very large spread in the UV to optical flux density arising
  from a large spread in the attenuation.  The colour scale in the
  upper image shows the 52 SEDs ranked by their characteristic dust
  temperature. These illustrate the wide variety in both the restframe
  UV/optical/near-infrared and mid-infrared characteristics of SMGs
  with very similar far-infrared luminosities.}
\label{fig:seds_edc}
}\end{center}
\end{figure}

Using our sample of spectroscopically confirmed SMGs with extensive
UV-to-radio photometry, we employ the {\sc magphys} SED fitting code
from \citep[see][]{DaCunha15} to fit the UV-to-radio emission on a
galaxy-by-galaxy basis to estimate the dust reddening, far-infrared
luminosity and infer the stellar mass for each SMG.  Estimates of
these parameters have been made using photometric redshifts, but the
addition of spectroscopic redshifts removes some of the degeneracies
between photometric redshift, reddening and star-formation histories,
to allow more precise estimates to be made.  The
UV--mid-infrared photometry for the ALESS SMGs is given in
\citet{Simpson14}, whilst the (deblended)
\emph{Herschel}\,/\,SPIRE+PACS, ALMA and radio photometry are given in
\citet{Swinbank14} \citep[see also][]{DaCunha15}.  For each SMG, we
use {\sc magphys} to fit the photometry at the spectroscopic redshift,
and we show the best-fit SEDs (normalised by their 8--1000$\mu$m
luminosities) in Fig.~\ref{fig:seds_edc}
\footnote{The template SEDs are available from: \url{http://astro.dur.ac.uk/$\sim$ams/zLESS/}}.
These normalised,
rest-frame SEDs demonstrate a large range in the UV- to optical-flux
density which is driven by a large spread in the dust
attenuation.  Indeed, the estimated  extinction varies from
A$_{\rm V}\sim$\,0.5--7 magnitudes between SMGs \citep[see also][]{DaCunha15}.

%
%

%
%

\begin{figure}
  \begin{center}
      \psfig{figure=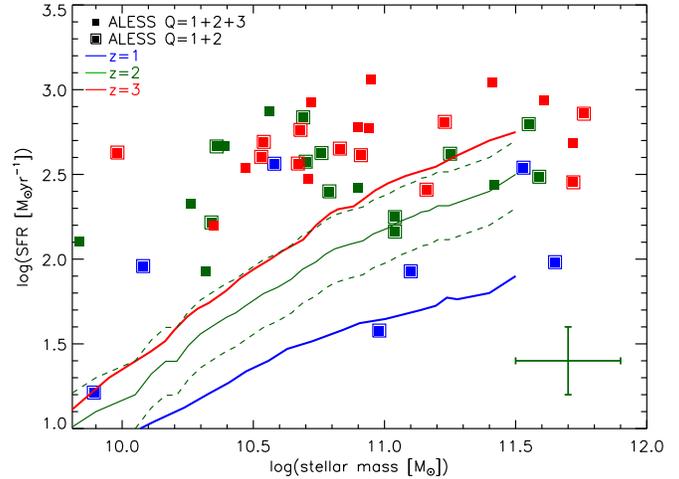,width=2.7in,angle=90}
      \caption{The stellar mass--star-formation rate plane for ALESS
        SMGs with spectroscopic redshifts compared to the so-called
        ``main-sequence'' of star-forming star-forming galaxies at
        $z=$\,1, 2 \& 3.  We identify the ALESS SMGs with the best
        spectroscopic redshifts (Q\,=\,1 \& 2) and the points are
        colour coded by their spectroscopic redshift.  Taken at face
        value plot suggests that at $z\sim$\,1--3 SMGs have sSFRs
        which lie between the ``main-sequence'' and an order of
        magnitude higher sSFR and on-average their sSFR are a factor
        of $\sim$\,5\,$\times$ higher than the bulk of the
        star-forming population at their stellar mass.  However, we
        caution that the stellar masses of these highly obscured and
        strongly star-forming galaxies are systematically uncertain
        \citep{Hainline11}.  We illustrate the expected conservative
        uncertainties for the measurements by the error bars plotted
        in the lower-right of the panel and stress that it is possible
        that the SMGs could be moved systematically by comparable
        amounts on this figure.  }
      \label{fig:ssfr}
  \end{center}
\end{figure}

From the sample, we derive a median extinction of A$_{\rm
  V}$\,=\,1.9\,$\pm$\,0.2 and far-infrared luminosity of $L_{\rm
  FIR}$\,=(3.2\,$\pm$\,0.4)\,$\times$\,10$^{12}$\,L$_{\odot}$, both of
which are consistent with previous estimates (for the same sample)
derived using photometric redshifts (A$_{\rm V}$\,=\,1.7\,$\pm$\,0.2
and $L_{\rm
  FIR}$\,=\,(3.5\,$\pm$\,0.4)\,$\times$\,10$^{12}$\,L$_{\odot}$
respectively from Simpson et al.\ (2014)).  In addition, {\sc magphys}
also returns estimates of the stellar masses (solving for the star
formation histories and ages) and we derive a median stellar mass for
our 52 SMGs with spectroscopic redshifts of
$M_\star$\,=\,(6\,$\pm$\,1)\,$\times$\,10$^{10}$\,M$_{\odot}$, in
agreement with previous estimates for this sample using photometric
redshifts and simple assumptions about the star formation histories by
\citet{Simpson14}, see also \citet{DaCunha15}.  This is also
consistent with the stellar masses estimates for the radio-identified
submillimetre sources in the \citet{Chapman05} sample
($M_\star\sim$\,7\,$\times$\,10$^{10}$\,M$_{\odot}$;
\citealt{Hainline11}).  In Fig.~\ref{fig:ssfr} we plot the ALESS SMGs
with spectroscopic redshifts on the stellar mass--star-formation rate
plane.  For comparison, we overlay the trends proposed for the
so-called ``main-sequence'' of star-forming galaxies at $z=$\,1, 2 \&
3 and compare these to the SMGs in the same redshift slices.  From
this plot, it is clear that the SMGs in our sample lie (on average) a
factor $\sim$\,5 above the so-called ``main-sequence'' at all three
redshifts, with a median specific star-formation rates (sSFR) of
sSFR\,=\,(6\,$\pm$\,1)\,$\times$\,10$^{-9}$\,yr$^{-1}$ \citep[see also
  e.g.\ ][]{Magnelli12,Simpson14}.

\subsection{Velocity offsets between emission\,/\,absorption lines}
\label{sec:veloff}

Rest-frame UV\/\,optical spectroscopic analysis of high-redshift, 
star-forming galaxies have shown that redshifts derived from UV ISM
absorption lines typically display systematic blue-shifted offsets
from the systemic (nebular) redshifts
\citep[e.g.][]{Erb06c,Steidel10,Martin12}, whilst redshifts
determined from Ly$\alpha$ emission often show a systematic offset
redward of the systemic.  These velocity offsets are a consequence of
large scale outflows \citep[e.g.][]{Pettini02b,Steidel10}, where the
outflows material between the galaxy and the observer absorbs the UV
and scatter Ly$\alpha$ photons from the receeding outflow, redshifting
them with respect to the neutral medium within the galaxies.  For some
of the ALESS SMGs we are able to determine nebular, UV ISM and
Ly$\alpha$ redshifts, allowing us to compare to the results for other
star-forming populations.

\begin{figure}
\begin{center}{
\psfig{figure=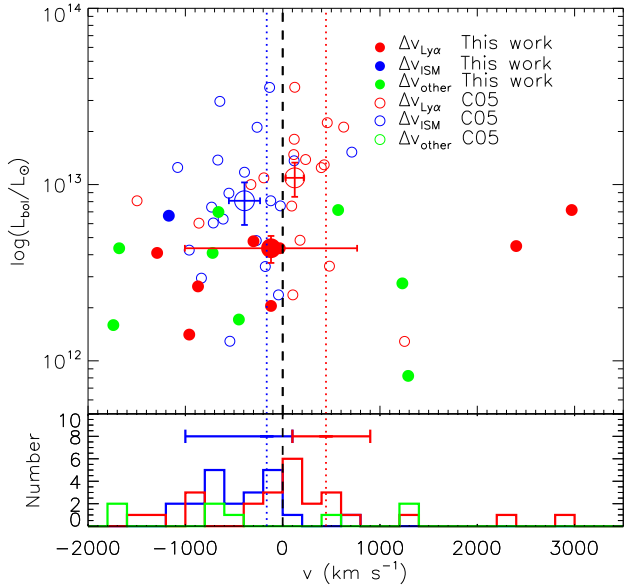,width=3.5in}
\caption{{\it Top:} Velocity offsets of the UV ISM absorption lines
  and Ly$\alpha$ from the systemic redshifts (marked by the dashed
  line) versus bolometric luminosity (L$_{8-1000\mu \rm m}$) for all
  ALESS SMGs and the radio-identified submillimetre sources from
  Chapman et al.\ (2005), where appropriate lines are detected.  The
  median of each sample is marked by a larger symbol. The red and blue
  dotted lines represent the mean of the distributions of Ly$\alpha$
  and ISM velocity offsets respectively from the $z=$\,2--3 LBG study
  from \citep{Steidel10} and the full range are shown as error bars on
  the bottom figure.  We show a representative error bar for our data
  derived from the median error on the bolometric luminosity and we
  estimate a typical redshift measurement error of
  $\sim$\,100\,km\,s$^{-1}$ from fitting the spectral lines.  The
  green points indicate offsets measured between lines which can be
  either nebular or ISM lines and are frequently strongly influenced
  by winds, such as C{\sc iv}\,$\lambda$1549, N{\sc v}\,$\lambda$1240,
  C{\sc iii}]\,$\lambda$1909, Mg{\sc ii}\,$\lambda$2800 and He{\sc
  ii}. Note that the far-infrared luminosities for the Chapman et
al.\ (2005) sources are derived from their radio fluxes and may be
overestimated.  {\it Bottom:} Histograms of the distributions of
velocity offsets for Ly$\alpha$ (red), UV ISM lines (blue) and other
lines (green).  The histograms include the SMGs from ALESS and the
radio-identified submillimetre sources in \citet{Chapman05}, and
demonstrate that Ly$\alpha$ and the UV ISM lines in SMGs do indeed
respectively peak redward and blueward of the systemic velocity, as
expected if these systems are driving outflows and winds.  }
\label{fig:zoffsets}
}\end{center}
\end{figure}

In Table~2 we summarise the lines detected for each ALESS SMG and the
redshift associated with fitting to each line.  We show the velocity
offsets between the Ly$\alpha$, UV ISM and nebular emission lines in
Fig.~\ref{fig:zoffsets}.  We also overlay the velocity offsets for the
radio-identified counterparts to submillimetre sources studied by
\citet{Chapman05}.  Although the same trend is seen in the SMGs and
LBGs (Ly$\alpha$ is redshifted and the UV ISM lines are blueshifted
with respect to the systemic redshift), the SMGs display significantly
more scatter, with velocity offsets ranging between $\sim -1100$ to
$+700$\,km\,s$^{-1}$ for the UV ISM-derived redshifts and between
$\sim -1500$ to $+1200$\,km\,s$^{-1}$ for the Ly$\alpha$-derived
redshifts, as compared to $-600$ to $+100$\,km\,s$^{-1}$ and
$\sim+100$ to $+900$\,km\,s$^{-1}$ respectively for the LBGs in
\citet{Steidel10}.  The wide variation in the velocity offsets may be
due to a spread in the viewing angle of the winds or the presence of
multiple components (\citealt{Chen15} suggest that most SMGs are major
mergers and so the spectra may have contributions from merging
components), or the diversity of conditions within these SMGs, in
particular with regard to the strength of large-scale winds.  Since
the wind must be accelerated by star formation or AGN activity, in
Fig.~\ref{fig:zoffsets} we plot the velocity offsets between lines as
a function of bolometric luminosity (we note that only two SMGs in our
sample are X-ray AGN; \citealt{Wang13} and neither of these show
Ly$\alpha$ and UV ISM lines with extreme offsets from the systemic
redshift).  Although there is significant scatter, within the ALESS
sample the SMGs with lower bolometric luminosity tend to have wind
velocities that are lower than those of the highest luminosity
sources.

%
%

We note that the outliers in Fig.~\ref{fig:zoffsets} are ALESS\,088.5
and ALESS\,049.1, with Ly$\alpha$ offset from the systemic  by
$>$\,2000\,km\,s$^{-1}$.  For both ALESS\,088.5 and ALESS\,049.1 the
only line available to determine a nebular\,/\,systemic velocity was
He{\sc ii}\,$\lambda$1640, which, as we described previously can
originate from the stellar winds from Wolf-Rayet stars, making it less
reliable as a systemic velocity tracer than the typical nebular lines
(e.g.\ H$\alpha$). It is important to note that the nebular lines such
as H$\alpha$, [O{\sc iii}] and [O{\sc ii}] may also be influenced by
winds, however this is more typically observed as line broadening as
opposed to a centroid shifting.

%
%

\subsection{Environments}
\label{sec:smg_env}

One of the key benefits from obtaining spectroscopic redshifts for
SMGs is the capability they provide to study both the small- and
larger-scale environments of these sources.  Hence, we next use our
spectroscopic redshift sample to search for physical associations
between SMGs and between SMGs and other galaxy populations within the
field.  Various studies have investigated the environments of SMGs and
suggested that at least some SMGs reside within overdense environments
\citep[e.g.][]{Chapman01,Blain04a,Chapman09,Daddi09,Capak11,Walter12,Ivison13,Decarli14,Smolcic16}.
For example, \citet{Blain04a} \citep[see also][]{Chapman09} identified
an over-density of six SMGs and two radio galaxies at $z=$\,1.99
within 1200\,km\,s$^{-1}$ of each other in the GOODS-N field.
Clustering analysis has also suggested that SMGs cluster on scales of
5--10\,$h^{-1}$\,Mpc$^{-1}$, while pair counting suggests SMGs have
properties consistent with them evolving into the passive red galaxies
at $z\sim$\,1, and subsequently the members of rich galaxy groups or
clusters at $z\sim$\,0
\citep[e.g.][]{Blain04a,Hickox12,Chen16,Wilkinson16}.

A potentially related result was found by \citet{Karim13}, who
demonstrated that single dish submillimetre sources suffer significant
``multiplicity'' \citep[see also][]{Simpson15b}, with $>$\,35\% of the
single dish sources resolved into multiple SMGs (where an SMG is a
far-infrared bright galaxy with a 870\,$\mu$m flux brighter than
1\,mJy).  \citet{Simpson15b} also showed that the number density of
$S_{\rm 870}\gsim$\,2\,mJy SMGs in ALMA maps of bright single-dish
submillimetre sources is $\sim$\,80 times higher than that derived
from blank-field counts.  After taking into account the observational
biases in their sample, they proposed that an over-abundance of faint
SMGs of this magnitude is inconsistent with line-of-sight projections
dominating multiplicity in the brightest SMGs, and strongly suggests
that a significant proportion of these high-redshift ULIRGs are likely
to be physically associated.  These SMGs are typically separated by
$\sim $\,6$''$ which corresponds to $\sim$\,40--50\,kpc if they lie at
the same redshift.

With our survey, we can use a simple approach and exploit the
spectroscopic redshifts to search for associations and overdensities
in the ALESS SMG population.  First, we search for physical
associations between SMGs in the same ALMA map (i.e.\ within
$\sim$\,18$''$) where the SMGs lie within 2000\,km\,s$^{-1}$ (although
an offset of 2000\,km\,s$^{-1}$ is larger than the typical velocity
dispersion of rich clusters, even at $z\sim$\,0, we broaden our search
window to account for potential outflow-driven shifts in the spectral
features used to derive the redshifts of many of the SMGs (see
\S\ref{sec:veloff}).  Unfortunately, there are only three ALESS maps
in which we were able to determine a reliable spectroscopic redshift
for two or more of the SMGs (ALESS\,017.1, 017.2; 075.1, 075.2; 088.1,
088.2, 088.5 and 088.11), and in none of these maps do we find any
small-scale clustering of SMGs along the line of sight, the range of
redshift offsets between these (previously blended) components is
$\Delta z=$\,0.06--1.25.  Only in ALESS\,067 do we have indirect
evidence for an interacting pair of SMGs (ALESS\,067.1 and
ALESS\,067.2) based on the morphology of the sources in \emph{HST}
imaging \citep{Chen15}.

Next, we search for physical associated between SMGs across the whole
ECDFS field (i.e.\ between the ALMA maps).  We identify seven pairs of
SMGs within 2000\,km\,s$^{-1}$ of each other, with ALESS\,075.2,
ALESS\,088.5 and ALESS\,102.1 also appearing as a triple
``association''. These pairs/triples of SMGs have an average offset of
$\sim$\,4\,Mpc in the plane of the sky (with a range of
$\sim$\,2--15\,Mpc).  On these scales, the pairs (or triples) may lie
within the same large-scale structure but are unlikely to lie within
the same dark matter halos.

To determine whether these potential ``associations'' correlate with
redshift peaks in other galaxy populations we compare the
spectroscopic redshift distribution of the ALESS SMGs with that of the
infill targets from our survey, as well as other archival surveys.
Most of the spectroscopic redshifts for the other galaxy populations
were taken from an updated version of the redshift compilation in
\citet{Luo11} listing $>$\,15,000 spectroscopic redshifts for galaxies
in the ECDFS with a median redshift of $z\sim$\,0.67 and an
inter-quartile range of $z=$\,0.3--1.0,\footnote{
  http://www.eso.org/sci/activities/garching/projects/goods/\\MASTERCAT\_v3.0.dat
  which includes redshifts from
  \citet{Cristiani00,Croom01,Bunker03,Dickinson04,Stanway04a,Stanway04b,Strolger04,Szokoly04,Vanderwel04,Lefevre05,Doherty05,
    Mignoli05,Ravikumar07,Vanzella08,Popesso09,Balestra10,Coppin10,Silverman10,Kurk13};
  and redshifts also taken from
  \citet{Kriek08,Boutsia09,Taylor09,Treister09,Wuyts09,Casey11b,Xia11,Bonzini12,Cooper12,Coppin12,Iwasawa12,Mao12,Lefevre13,Georgantopoulos13,DeBreuck14,Williams14}
  and the 2dF Galaxy Redshift Survey \citep{colless03}}.  From this
catalog, we select only secure redshifts and remove duplicates (we
also remove cases in which two secure but differing redshifts are
given from two different references).

\begin{figure*}
\begin{center}{
\psfig{figure=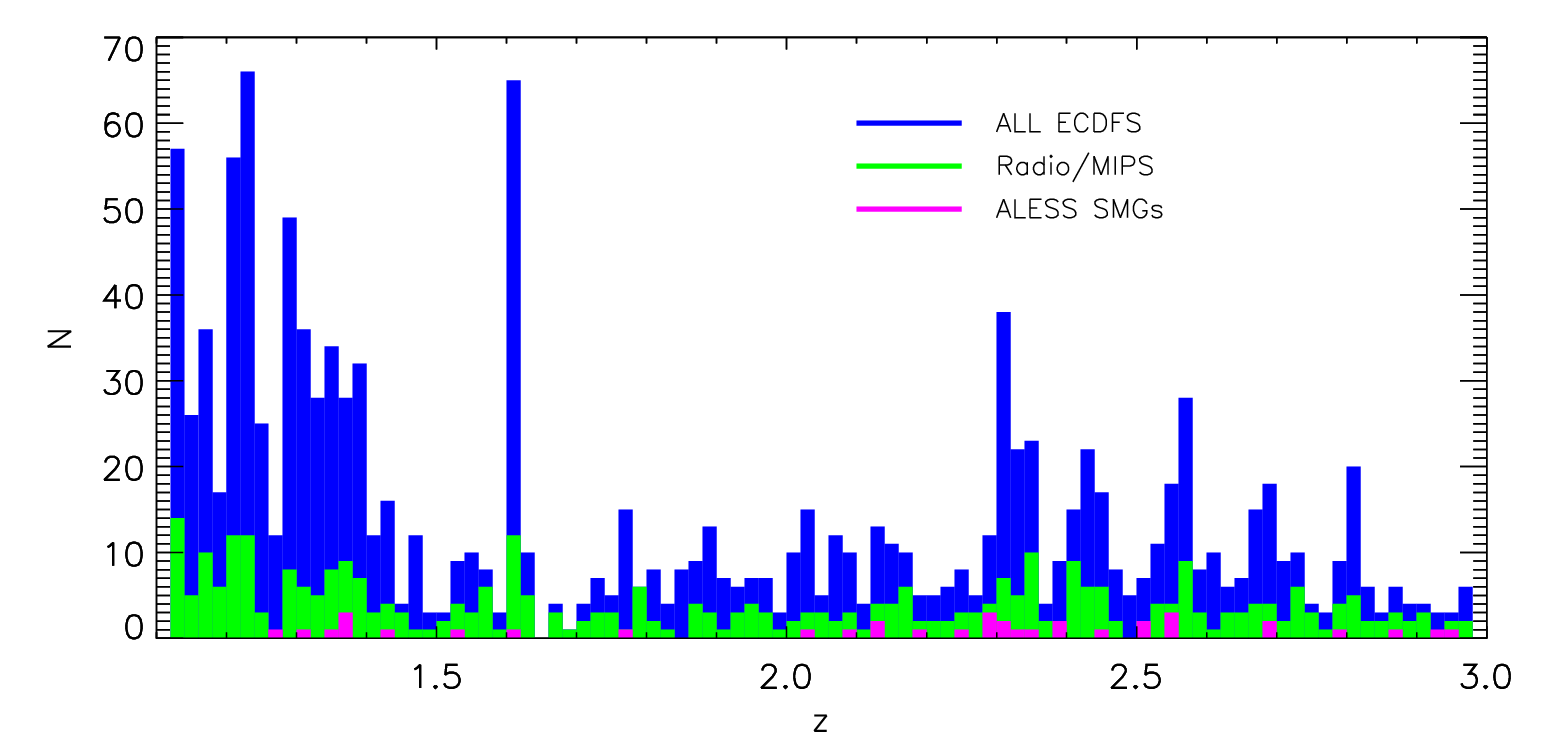,width=7in}
\psfig{figure=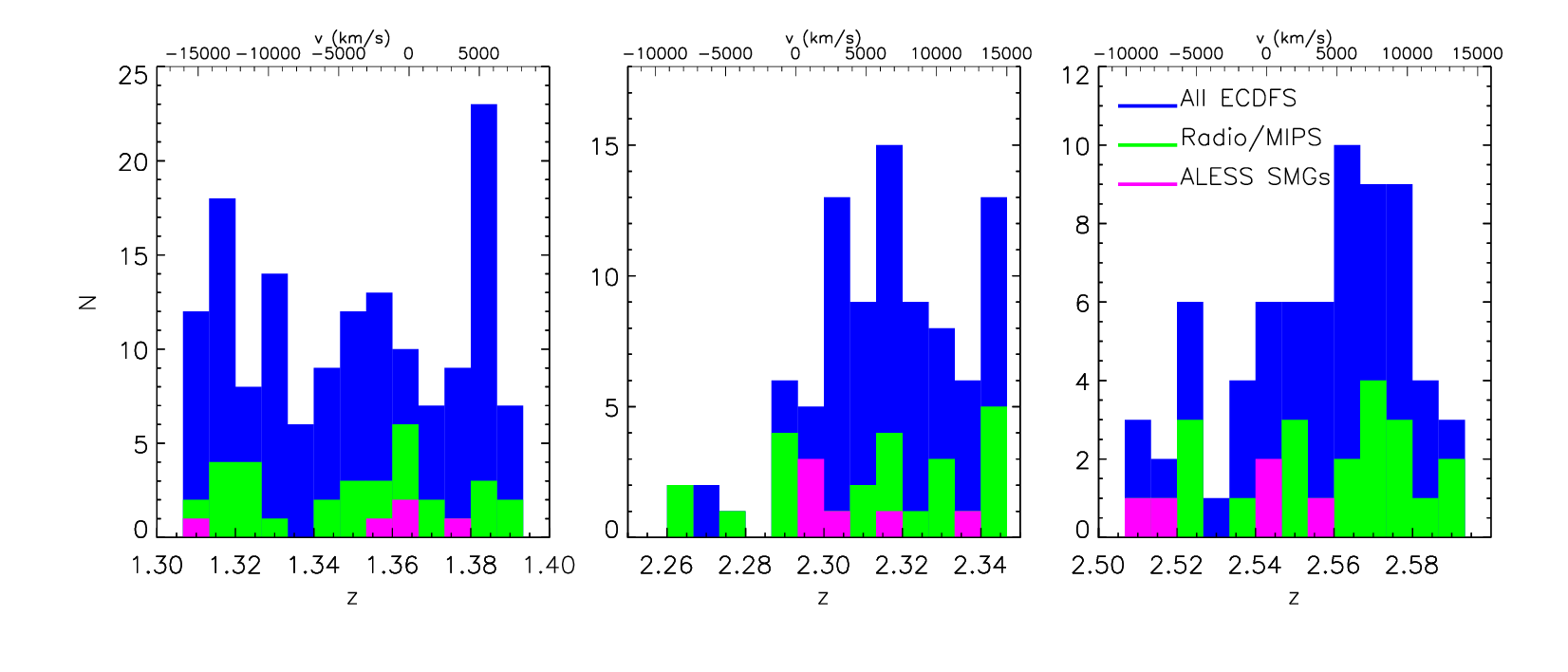,width=7in}
\caption{{\it Top:} The spectroscopic redshift distribution of SMGs
  (Q\,=\,1, 2 \& 3) compared to the less luminous galaxy populations
  in the field. The latter is based on the catalogue compiled by Luo
  et al.\ (2011) with the addition of recent redshifts from the full
  FORS2/VIMOS survey (Table~2 \& 3) and from \citet{Williams14}. We
  plot all the galaxies in the ECDFS for which we have spectroscopic
  redshifts (including the SMGs), we also plot the distributions for
  just the radio/MIPS sources, as well as the SMGs. We see little
  correlation between the peaks in the SMG redshift distribution and
  the general galaxy distribution.  The binning is 6000\,km\,s$^{-1}$
  in all panels.  {\it Bottom:} Expanded views of the redshift
  distribution around the associations of the ALESS SMG compared to
  the overall galaxy redshift distribution.  We find a maximum of
  three SMGs in our adopted 2000\,km\,s$^{-1}$ window, in addition to
  three pairs of SMGs.  The pairs/triples in the SMG population do not
  obviously coincide with overdensities in the less-active galaxy
  populations across the field.  The colour coding is the same as
  Fig.~\ref{fig:alessfield} and the top axis indicates velocity
  relative to the redshift of the pair/triple.  }
\label{fig:alessfield}
}\end{center}
\end{figure*}

In Fig.~\ref{fig:alessfield} we plot the spectroscopic redshift
distribution of the ALESS SMGs, together with the field population.
In those cases where $\geq$\,2 SMGs lie within 2000\,km\,s$^{-1}$,
these associations do not often statistically coincide with
significant over-densities in the background galaxy population,
although the two SMGs at $z\sim$\,1.36 are coincident with a slight
peak in the radio\,/\,MIPS sources at that redshift.

Finally, returning to Fig.~\ref{fig:hubble} we have highlighted there
the ten SMGs that are members of pairs (or triples) with spectroscopic
redshift offsets between components of $\le$\,2000\,km\,s$^{-1}$.  The
median apparent magnitude at 4.5\,$\mu$m for these ten SMGs is
$m_{4.5\mu \rm m}=$\,20.4$^{+0.7}_{-0.6}$ as compared to a median of
$m_{4.5\mu \rm m}=$\,21.1$^{+0.1}_{-0.4}$ for the 42 ALESS SMGs in the
parent spectroscopic sample which are not in identified
``associations''.  We conclude that there is no evidence in the
current sample that the SMGs in ``associations'' are any brighter (and
thus potentially more massive) than those not in ``associations''.

%
%

\section{Conclusions}
\label{sec:conc}

In this work we present the results from a redshift survey of
ALMA-identified SMGs.  Our main conclusions are:

\begin{itemize}
\item The redshift distribution for ALESS SMGs with spectroscopic
  redshifts is centered at $z=$\,2.4\,$\pm$\,0.1, but with a full
  range of $z=$\,0.7--5.0 and an interquartile range of
  $z=$\,2.1--3.0.  This is consistent with the photometric redshift
  distribution for these sources, and the median is consistent with
  previous estimates based on the radio-identified counterparts to
  submillimetre sources \citep{Chapman05}.  However, since we do not
  rely on a radio selection, our sample is not biased against higher
  redshift SMGs and indeed, 23\% of the ALESS SMGs with spectroscopic
  redshifts lie at $z>$\,3.
\item We identify velocity offsets up to $\sim$\,3000\,km\,s$^{-1}$
  between the redshifts measured from nebular emission lines
  (i.e.\ H$\alpha$, [O{\sc iii}], H$\beta$ and [O{\sc ii}]) and those
  measured from Ly$\alpha$ or UV ISM absorption lines.  We conclude
  that it is likely that the extreme SFRs within the SMGs (typically
  $\sim$\,300\,$\pm$\,30\,M$_{\odot}$\,yr$^{-1}$) are driving strong
  galaxy-scale outflows in many of these systems.
\item Since many of our spectra of SMGs are too faint to exhibit any
  obvious emission or absorption features (continuum is only detected
  in $\sim$\,50\% of the sources), we produce composite spectra over
  various wavelength ranges to search for weaker features in the
  ``typical'' ALESS SMG optical-to-near infrared spectrum.  At
  rest-frame 1000--2000{\AA} we see strong, asymmetric Ly$\alpha$
  emission and blueshifted Si{\sc ii} and potentially Si{\sc iv}
  absorption suggestive of strong stellar winds.  Our composite
  spectrum at rest-frame 3400--4400{\AA} shows a Balmer break,
  indicative of on-going star formation.  Comparing our composite to
  spectral models we suggest that it is most consistent with a young
  starburst with an age of $\sim$\,10\,Myr.
\item We use our precise spectroscopic redshifts to reduce the
  uncertainties when modelling the SEDs of our SMGs using {\sc
    magphys} and find a large spread in the dust attenuation (A$_{\rm
    V}\sim$\,0.5--7 magnitudes) with a median A$_{\rm
    V}$\,=\,1.9\,$\pm$\,0.2.  We also derive a median stellar mass of
  $M_\star$\,=\,(6\,$\pm$\,1)\,$\times$\,10$^{10}$\,M$_{\odot}$ and by
  combining with our estimates of their star-formation rates, we show
  that SMGs lie (on average) $\sim$\,5 times above the so-called
  ``main-sequence'' at $z\sim$\,1--3.  We provide this library of
  template SEDs for 52 SMGS with precise redshifts and well-sampled
  photometry as a resource for future studies of SMGs.
\end{itemize}

This work has highlighted the challenges of measuring spectroscopic
redshifts at optical-to-near infrared wavelengths for dusty
star-forming galaxies identified by ALMA, and thus demonstrates the
importance of alternative methods of measuring redshifts such as
mid-infrared spectroscopy \citep[e.g.][]{Valiante07} and the
increasing importance of blind submillimetre\,/\,millimetre spectral
searches with ALMA \citep[e.g.][]{Weiss13}.

Neverthless, we find that the SMG population is a diverse population
of dusty galaxies most common at $z\sim$\,2.4, with  evidence of
energetic outflows which are likely to be predominantly driven by star
formation, although some may have a contribution from AGN.  The main goal of this
study was to provide redshifts for subsequent studies such as CO gas
 or further detailed integral field unit (IFU) follow-up
observations. Such studies will allow us to separate out the relative
contributions of star formation and AGN, to probe the conditions
within the star-forming gas to better understand this extreme and
diverse population of galaxies.

\section*{Acknowledgments}

We acknowledge the ESO programmes 183.A-0666 and 090.A-0927(A).  The
ALMA observations were carried out under programme 2011.0.00294.S.
ALRD acknowledges an STFC studentship (ST/F007299/1) and an STFC STEP
award. AMS gratefully acknowledges an STFC Advanced Fellowship through
grant ST/H005234/1, STFC grant ST/L00075X/1 and the Leverhume
foundation.  IRS acknowledges support from STFC, a Leverhulme
Fellowship, the ERC Advanced Investigator programme DUSTYGAL 321334
and a Royal Society/Wolfson Merit Award.  WNB acknowledges STScI grant
HST-GO-12866.01-A.  CMC acknowledges support from a McCue Fellowship
at the University of California, Irvine’s Center for Cosmology and the
University of Texas at Austin’s College of Natural Science.  JLW is
supported by a European Union COFUND/Durham Junior Research Fellowship
under EU grant agreement number 267209.  AK acknowledges support by
the Collaborative Research Council 956, sub-project A1, funded by the
Deutsche Forschungsgemeinschaft (DFG).  ALMA is a partnership of ESO
(representing its member states), NSF (USA) and NINS (Japan), together
with NRC (Canada) and NSC and ASIAA (Taiwan), in cooperation with the
Republic of Chile. The Joint ALMA Observatory is operated by ESO,
AUI/NRAO and NAOJ.

\appendix

\section{ALESS SMGs with literature redshifts}
\label{sec:litred}

The following sources are ALESS SMGs with previously measured
spectroscopic redshifts:

\begin{enumerate}
\item ALESS\,018.1: is listed as ID\,66 in \cite{Casey11b}, with a
  redshift of $z=$\,2.252 derived from an H$\alpha$ detection with the
  Infrared Spectrometer And Array Camera (ISAAC) on the VLT;
\item ALESS\,057.1: is listed as ID\,112a in \cite{Szokoly04} with a
  redshift of $z=$\,2.940 derived from detections of He{\sc ii}, O{\sc vi} and N{\sc v} with
  FORS1\,/\,FORS2.  It is classed as a QSO with strong
  high-ionisation emission lines;
\item ALESS\,067.1: is listed as ECDFS-45 in \cite{Kriek08} at 
$z=$\,2.122, derived from emission lines in the
  near-infrared spectrum observed with GNIRS;
\item ALESS\,073.1: is listed as GDS J033229.29$-$275619.5 in the
  \citet{Vanzella08} compilation of 1019 spectroscopic redshifts for
  GOODS\,/\,CDFS. The redshift of $z=$\,4.762 was determined via the
  detection of Ly$\alpha$ and N{\sc v} using FORS2.
\item ALESS\,098.1: is identified as ID\,J033129 in \cite{Casey11b}. The
  redshift, $z=$\,1.4982 is derived through a tentative detection of
  H$\alpha$, however, it is also spectroscopically-identified in the
  restframe UV in the same paper and therefore it is given a ``secure''
  status. This redshift is, however, in disagreement with our Q\,=\,1
  redshift of $z=$\,1.3735 derived from fitting to an [O{\sc ii}]
  line in the FORS2 observations, with a tentative detection of
  H$\alpha$ at the same redshift under a sky line in the XSHOOTER near-infrared
  spectrum. We use our redshift in the analysis in this work;
\item ALESS\,122.1: is listed as radio ID\,149 in \cite{Bonzini12}. The
  redshift of $z=$\,2.03 is determined from UV ISM absorption
  features observed with VIMOS.
\end{enumerate}

\section{Notable Individual Sources}
\label{sec:indiv}

Since we have a wealth of spectroscopic data we can utilise the
spectra not only for the purpose of determining redshifts but also to
search for diagnostic features indicative of AGN activity, star
formation, strong stellar winds etc.  Here we highlight and discuss
some of the most notable, high signal-to-noise spectra. \\

{\bf ALESS\,057.1:} This SMG hosts a luminous AGN which is detected in
X-rays \citep{Wang13}.  The VIMOS spectrum (Fig.~\ref{fig:spec1})
exhibits strong, broad, symmetric Ly$\alpha$ emission, broad N{\sc v}
and C{\sc iv} emission (FWHM\,$\sim$\,3700\,km\,s$^{-1}$) which is
significantly blue-shifted ($\sim$\,1600\,km\,s$^{-1}$) with respect
to both He{\sc ii} and Ly$\alpha$ (which have velocities that are
consistent within measurements errors).  The C{\sc iv} emission line
also displays a P-Cygni profile.

\noindent{\bf ALESS\,066.1:} This SMG is listed as an X-ray AGN at
$z=$\,1.310 in \cite{Wang13}.  However, our observations reveal the
optical/near-infrared photometry and X-ray emission are dominated by a
foreground QSO at $z=$\,1.310 but our near-infrared spectroscopy with
MOSFIRE identifies an emission line in $K$-band slightly to the north
of the QSO. At $\lambda$\,=\,2.333\,$\mu$m this line corresponds to
H$\alpha$ at $z=$\,2.5542.  Careful analysis of the ALMA and optical
imaging reveals that the SMG is indeed $\lsim$\,1$''$ north of the QSO
and hence is likely to be lensed by the foreground QSO.

\noindent{\bf ALESS\,073.1:} This SMG also hosts a luminous X-ray AGN
\citep{Vanzella08,Coppin09,DeBreuck14,Wang13} and the spectrum
(Fig.~\ref{fig:spec1}) shows strong, broad N{\sc v} with a
FWHM\,$\sim$\,3000\,km\,s$^{-1}$ as compared to a relatively narrow
and weak Ly$\alpha$ (FWHM\,$\sim$\,700\,km\,s$^{-1}$).
 
\noindent{\bf ALESS\,075.1:} We have excellent spectroscopic coverage
of this SMG and have strong detections of [O{\sc ii}], [O{\sc
    iii}]$\lambda$4959, 5007, H$\beta$ and H$\alpha$ with XSHOOTER.
The H$\alpha$ detection is narrow with
FWHM\,$\sim$\,160\,km\,s$^{-1}$.  The [O{\sc iii}] emission is not fit
well with a single Gaussian as it is an asymmetric line with a red
wing, possibly indicating an outflow \citep[e.g.][]{Alexander10}.
Given the high [O{\sc iii}] luminosity and the lack of an X-ray
detection, this outflow may be accelerated by an obscured AGN
(i.e.\ outflows in high-redshift ULIRGs hosting AGN activity;
\citealt{Harrison12}).

\begin{figure*}
\begin{center}{
\psfig{figure=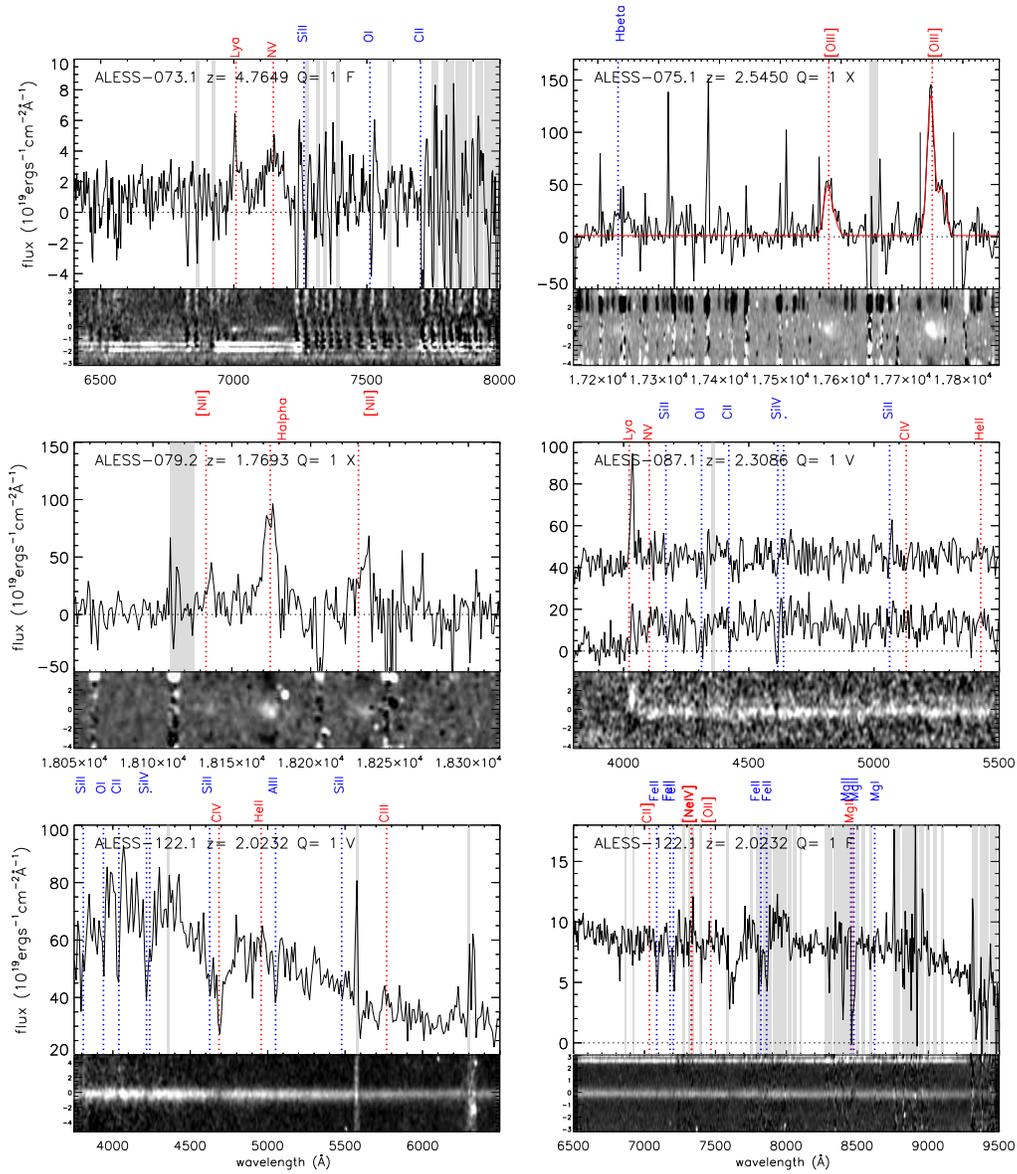,width=5.5in}
\caption{Some of the most notable spectra of SMGs in the sample,
  featuring evidence of winds, AGN activity, and multiple
  components. The sky subtraction is poor in some of the spectra and
  is a particular problem in the near-infrared and in the FORS2
  spectrum of ALESS\,073.1. The main skylines have been highlighted in
  grey. }
\label{fig:interest}
}\end{center}
\end{figure*}

\noindent{\bf ALESS\,079.2:} This SMG has strong detections of
H$\alpha$ and [N{\sc ii}] with XSHOOTER.  The one- and two-dimensional
spectra show structured emission (see Fig.~\ref{fig:interest}).  In
the one-dimensional spectrum the H$\alpha$ and [N{\sc ii}] lines are
truncated at their red end and appear to be more extended towards
lower velocities.  The flux ratio of [N{\sc
    ii}]$\lambda$6583/H$\alpha$ is consistent with the ionising
radiation arising from H{\sc ii} regions as opposed to an AGN.

\noindent{\bf ALESS\,087.1:} Strong rest-frame UV continuum is
detected in this SMG with ISM absorption lines, with reshifts
consistent with the Ly$\alpha$ emission line. However, the Ly$\alpha$
is significantly offset northwards of the continuum in the
two-dimensional spectrum.  We therefore extract two spectra in
Fig.~\ref{fig:interest} taken from the position of the Ly$\alpha$ and
the continuum.  The Ly$\alpha$ profile is marginally asymmetric with a
truncated blue edge.  The continuum spectrum shows an obvious break
and relatively strong Si{\sc iv} absorption.  Unfortunately, there is
very poor photometric coverage of this SMG (3.6--8\,$\mu$m only) so we
are unable to say whether the offset Ly$\alpha$ is due to a close
companion or an interaction with another system, or a less-obscured
part of a single galaxy.

\noindent{\bf ALESS\,122.1:} This SMG has very blue continuum with
strong UV ISM absorption lines in both the FORS2 and VIMOS spectra
(Fig.~\ref{fig:interest}).  There is very strong, broad C{\sc iv}
absorption (FWHM of $>$\,7000\,km\,s$^{-1}$).  The C{\sc iv} exhibits
a strong, narrow component associated with interstellar absorption and
a very broad red component associated with stellar winds. The strength
of this redshifted component suggests the presence of a large number
of very massive stars ($>$\,30\,M$_{\odot}$;
\citealt{Leitherer95}). Models show that Si{\sc iv} is relatively weak
for a continuous star formation history but yields a strong P-Cygni
profile for bursty star formation.  Detection of a P-Cygni profile for
Si{\sc iv} is therefore a good indicator that the burst duration is
short relative to the age.  The Si{\sc iv} absorption feature is
unusually broad ($>$\,3000\,km\,s$^{-1}$). This is the blueshifted
wind absorption.  \cite{Swinbank14} determine L$_{\rm
  FIR}=(6.0\pm0.4)\times10^{12}$\,L$_{\odot}$ for this SMG which
implies a star-formation rate of
SFR\,$\sim1040\pm70$\,M$_{\odot}$yr$^{-1}$ (using
\citealt{Kennicutt98}) which is higher than typical ALESS SMGs,
SFR\,$\sim(310\pm30)$\,M$_{\odot}$yr$^{-1}$ \citep{Swinbank14}.  We
note that an AGN may also exhibit strong C{\sc iv} absorption and
given the very strong continuum and the large width of the C{\sc iv}
in this SMG, it is plausible that it may be a broad absorption line
(BAL) AGN.

%
%


\clearpage
%
%
\begin{sidewaystable}[h]
  \scriptsize
  \begin{tabular}{lllllllll}
    \multicolumn{9}{c}
                {\textit{Table 2: ALESS spectroscopic redshift catalog}}\\
    \hline\hline
      ALESS ID               & RA                 & DEC                 & z$_{spec}$    & Q$_{spec}$    & z$_{\rm phot}^a$            & M/S$^b$     & Instruments$^c$ & Notes  \\
                             & (J2000)            & (J2000)             &              &             &                           &             &             & \\ 
      \hline
      ALESS 001.1            &  53.310270         & -27.937366          & 4.9540       & 3           & $4.34^{+2.66}_{-1.43}$       & M           & GMX         & [O{\sc ii}] in M-$K$       \\
      ALESS 001.2            &  53.310059         & -27.936562          & ...          & 4           & $4.65^{+2.34}_{-1.02}$       & M           & FVX         & BLANK                      \\
      ALESS 001.3            &  53.309069         & -27.936759          & ...          & 4           & $2.85^{+0.20}_{-0.30}$       & M           & X           & BLANK                      \\
      ALESS 002.1            &  53.261188         & -27.945211          & 2.1913       & 3           & $1.96^{+0.27}_{-0.20}$       & M           & DV          & poss. C{\sc iii}] em in D  \\
      ALESS 002.2            &  53.262800         & -27.945252          & ...          & 4           & -                         & M           & D           & BLANK                      \\
      ALESS 003.1            &  53.339603         & -27.922304          & 4.2373       & 3           & $3.90^{+0.50}_{-0.59}$       & M           & FMV         & poss. Ly$\alpha$ em in F+V \\
      \textit{ALESS 003.2}   & \textit{53.342461} & \textit{-27.922486} & \textit{...} & \textit{4 } & {\it 1.44$^{+0.43}_{-0.38}$} & \textit{S}  & \textit{M}  & \textit{BLANK}             \\
      {\it ALESS 003.3}      &  {\it 53.336294}   & {\it -27.920555}    & {\it  ...}   & {\it 4}     & -                         & {\it S}     & {\it M}     & {\it BLANK}                \\
      {\it ALESS 003.4}      &  {\it 53.341644}   & {\it -27.919379}    & {\it ...}    & {\it 4}  & - & {\it S}  & {\it M} 	   & {\it BLANK} \\
      ALESS 005.1            &  52.870467         & -27.985840          & ...          & 4  & $2.86^{+0.05}_{-0.04}$ & M & DMX   & BLANK \\	    
      ALESS 006.1            &  53.237331         & -28.016856          & 2.3338 & 1   & $0.45^{+0.06}_{-0.04}$ & M & GX & cont. from bright sources above SMG; Ly$\alpha$ em ($z=2.3295$) and \\
                             &                    &                     &              & & & & & C{\sc iv} em ($z=2.3314$) in X-UVB; H$\alpha$ and [O{\sc iii}]5007 in G  ($z=2.3338$)\\
      ALESS 007.1            &  53.314242         & -27.756750          & 2.6923 & 1   & $2.50^{+0.12}_{-0.16}$ & M & DFXS  & strong cont.; $z$ from H$\alpha$ in X-NIR; He{\sc ii} in X-VIS ($z=2.6901$)\\
      {\it ALESS 007.2}      &  {\it 53.312522}   & {\it -27.758499}    & {\it  ...}   & {\it 4}  & - & {\it S}  & {\it D}        & {\it BLANK} \\
      ALESS 009.1            &  53.047244         & -27.869981          & ...          & 4  & $4.50^{+0.54}_{-2.33}$ & M & D      & BLANK \\			
      ALESS 010.1            &  53.079418         & -27.870781          & 0.7616       & 1 & $2.02^{+0.09}_{-0.09}$ & M & FV & [O{\sc ii}] in V; [O{\sc ii}] ($z=0.7613$), [O{\sc iii}]4959 ($z=0.7619$), \\
                             &                    &                     &              & & & & & H$\beta$ ($z=0.7617$) in F; $z$ is mean from [O{\sc ii}], [O{\sc iii}], H$\beta$, possible lens \\
      ALESS 011.1            &  53.057688         & -27.933403          & 2.6832 & 2   & $2.83^{+1.88}_{-0.50}$ & M & FV       & Ly$\alpha$ em in V, no cont. \\			
      ALESS 013.1            &  53.204132         & -27.714389          & ...          & 4  & $3.25^{+0.64}_{-0.46}$ & M & DG      & BLANK \\   
      ALESS 014.1            &  52.968716         & -28.055300          & ...          & 4  & $4.47^{+2.54}_{-0.88}$ & M & VX       & BLANK \\   
      ALESS 015.1            &  53.389034         & -27.991547          & ...          & 4  & $1.93^{+0.62}_{-0.33}$ & M & DFGVX & BLANK \\   
      {\it ALESS 015.2}      &  {\it 53.391876}   & {\it -27.991724}    & {\it  ...}   & {\it 4}  & - &  {\it S}  & {\it M}	    & {\it BLANK} \\				
      ALESS 015.3            &  53.389976         & -27.993176          & 3.4252       & 3  & - & M & DM	    & Ly$\alpha$ em ($z=3.4399$) and C{\sc iv} em ($z=3.4106$) in D \\
      {\it ALESS 015.6}      &  {\it 53.388192}   & {\it -27.995048}    & {\it  ...}   & {\it 4}  & - & {\it S}  & {\it M}        & {\it BLANK} \\
      ALESS 017.1            &  53.030410         & -27.855765          & 1.5397       & 1  & $1.51^{+0.10}_{-0.07}$ & M & DFMV  & strong cont.; $z$ from H$\alpha$ in M-$H$; Mg{\sc ii} abs in F ($z=1.5382$) \\
      {\it ALESS 017.2}      &  {\it 53.034437}   & {\it -27.855470}    & {\it 2.4431} & {\it 3}  & {\it 2.10$^{+0.65}_{-1.37}$} & {\it S}  & {\it M} & {\it poss. H$\alpha$ in M-K} \\
      {\it ALESS 017.3 }     &  {\it 53.030718}   & {\it -27.859423}    & {\it  ...}   & {\it 4}  & {\it 2.58$^{+0.16}_{-0.32}$} & {\it S}  & {\it D}	    & {\it  BLANK} \\	
      ALESS 018.1            &  53.020343         & -27.779927          & 2.2520$^d$   & 1  & $2.04^{+0.10}_{-0.06}$ & M & V         & cont. in V; archival $z$ from Casey+11 \\
      ALESS 019.1            &  53.034401         & -27.970609          & ...          & 4  & $2.41^{+0.17}_{-0.11}$ & M & FV       & BLANK \\
      {\it ALESS 020.1}      &  {\it 53.319834}   & {\it -28.004431}    & {\it ...}    & {\it 4}   & {\it 2.58$^{+0.16}_{-0.32}$} &{\it S}  & {\it DFV }   & {\it cont. in F} \\
      {\it ALESS 020.2}      &  {\it 53.317807}   & {\it -28.006470}    & {\it ...}    & {\it 4}   & - & {\it S}  & {\it D}        &	{\it BLANK} \\
      ALESS 022.1            &  52.945494         & -27.544250          & ...          & 4 & $1.88^{+0.18}_{-0.23}$ & M & FV       & cont. in F+V \\
      ALESS 023.1            &  53.050039         & -28.085128          & ...          & 4   & $4.99^{+2.01}_{-2.55}$ & M & V         & BLANK \\
      ALESS 025.1            &  52.986997         & -27.994259          &  2.8719      &  3  & $2.24^{+0.07}_{-0.17}$ & M & V        & Ly$\alpha$ + break, cont. \\
      ALESS 029.1            &  53.403749         & -27.969259          & 1.438 9      & 2   & $2.66^{+2.94}_{-0.76}$ & M & DGMV & H$\alpha$ in M-$H$ \\
      ALESS 031.1            &  52.957448         & -27.961322          & ...          & 4   & $2.89^{+1.80}_{-0.41}$ & M & FVX    & BLANK \\
      {\it ALESS 034.1}      &  {\it 53.074833}   & {\it -27.875910}    & {\it 2.5115} & {\it 2}   & {\it 1.87$^{+0.29}_{-0.32}$} & {\it S}  & {\it M} & {\it broad H$\alpha$ in M-$K$}\\
      ALESS 035.1            &  52.793776         & -27.620948          & ...          &  4   & - & M & V        & BLANK \\			
      ALESS 037.2            &  53.401514         & -27.896742          & 2.3824       &  3   & $4.87^{+0.22}_{-0.40}$ & M & M & H$\alpha$ ($z=2.3824$) and [S{\sc ii}] ($z=2.3831$) \\
      {\it ALESS 038.1}      &  {\it 53.295153}   & {\it -27.944501}    & {\it  ...}   &  {\it 4}   & {\it 2.47$^{+0.11}_{-0.05}$} & {\it S}  & {\it D}        & {\it strong cont.+emission lines from contaminating source} \\
      ALESS 039.1            &  52.937629         & -27.576871          & ...          &  4   & $2.44^{+0.17}_{-0.23}$ & M & X	      & poss. faint lines, no cont. \\
      ALESS 041.1            &  52.791959         & -27.876850          & 2.5460       &  2   & $2.75^{+4.25}_{-0.72}$ & M & FV & strong cont. in F+V; C{\sc iii}]1909 em ($z=2.5459$), \\
                             &                    &                     &              & & &  & & C{\sc ii}]2326 em ($z=2.5500$) in F; cont. break in V \\
      ALESS 041.3            &  52.792927         & -27.878001          & ...          & 4    & - & M & M        & weak cont. \\
      ALESS 043.1            &  53.277670         & -27.800677          & ...          & 4    & $1.71^{+0.20}_{-0.12}$ & M & DFV     & possible faint lines, no cont. \\
      {\it ALESS 043.3}      &  {\it 53.276120}   & {\it -27.798534}    & {\it ...}    & {\it 4}    & - & {\it S}  & {\it D}	      & {\it BLANK} \\
      ALESS 045.1            &  53.105255         & -27.875148          & ...          & 4    & $2.34^{+0.26}_{-0.67}$ & M & FV       & no cont.; poss. Ly$\alpha$ em $z=2.9690$ from V and C{\sc iv} $z=2.9867$ from F \\
      {\it ALESS 046.1}      & {\it  53.402937}   & {\it -27.547072}    & {\it ...}    & {\it 4}    & - & {\it S}  & {\it FV  }     & {\it  faint cont. in F} \\
      ALESS 049.1            &  52.852998         & -27.846406          & 2.9417       & 2   & $2.76^{+0.11}_{-0.14}$ & M  & DFV    & strong cont. in F + V; He{\sc ii} em ($z=2.9417$), C{\sc iv} em ($z=2.9436$), \\
      \hline
      \label{tab:zaless}
  \end{tabular}
\end{sidewaystable}

\clearpage

\begin{sidewaystable}[h]
  \scriptsize
    \begin{tabular}{lllllllll}
    \multicolumn{9}{c}%
                {\textit{Continued from previous page}}\\      \hline\hline
ALESS ID               & RA                 & DEC                 & z$_{spec}$    & Q$_{spec}$    & z$_{\rm phot}^a$            & M/S$^b$     & Instruments$^c$ & Notes  \\
                       & (J2000)            & (J2000)             &              &             &                           &             &             & \\ 
\hline
ALESS 049.2            &  52.851956         & -27.843914          & ...      & 4   & $1.47^{+0.07}_{-0.10}$ & M  & M	      & BLANK \\
ALESS 051.1            &  52.937754         & -27.740922          & 1.3638 & 3   & $1.22^{+0.03}_{-0.06}$ & M  & FV & strong cont. in F+V, [O{\sc ii}] ($z=1.3638$) and break $\sim8000$\AA \\
                       &                    &                     & & & & & & and poss. Mg{\sc ii} em ($z=1.3681$) in F   \\
ALESS 055.1            &  53.259242         & -27.676513          & 1.3564 & 2   & $2.05^{+0.15}_{-0.13}$ & M  & DF & strong cont. in F+D; Mg{\sc ii}em ($z=1.3556$) \\
                       &                    &                     & & & & & & and H+K abs. (Kabs. $z=1.3572$) in F \\
ALESS 055.2            &  53.258983         & -27.678148          & ...      & 4   & - & M  & D	      & BLANK \\
ALESS 057.1            &  52.966348         & -27.890850          &  2.9369$^d$ &  1  & $2.95^{+0.05}_{-0.10}$ & M  & FV       & cont. + Ly$\alpha$ em ($z=2.9387$), C{\sc iv} em ($z=2.9332$), \\
                       &                    &                     & & & & & & He{\sc ii} em ($z=2.9388$) in V \\  
ALESS 059.2            &  53.265897         & -27.738390          & ...      &  4  & $2.09^{+0.78}_{-0.29}$ & M  & X	      & BLANK \\   
ALESS 061.1            &  53.191128         & -28.006490          & 4.4190 &  1  & $6.52^{+0.36}_{-0.34}$ & M  & A 	      & ALMA [C{\sc ii}]158$\mu$m\\
{\it ALESS 062.1}      &  {\it 53.150677}   & {\it -27.580258}    & {\it ...}      &  {\it 4}  & - &  {\it S}  & {\it D}         & {\it BLANK} \\		
{\it ALESS 062.2}      &  {\it 53.152410}   & {\it -27.581619}    & {\it 1.3614} &  {\it 1}  & {\it 1.35$^{+0.08}_{-0.11}$} &{\it S}   & {\it DFV} & {\it [O{\sc ii}] in D+F.  [O{\sc ii}] doublet resolved in D. }\\
ALESS 063.1            &  53.285193         & -28.012179          & ...      & 4   & $1.87^{+0.10}_{-0.33}$ & M  & G         & poss. faint em lines \\
ALESS 065.1            &  53.217771         & -27.590630          & 4.4445 & 1   & - &  M  & AD       & z from ALMA [C{\sc ii}158]$\mu$m, Ly$\alpha$ \\ 
ALESS 066.1            &  53.383053         & -27.902645          & 2.5542 & 1   & $2.33^{+0.05}_{-0.04}$ & M  & FMV & H$\alpha$ and [N{\sc ii}] in M; lensed? \\	
ALESS 067.1            &  53.179981         & -27.920649          & 2.1230$^d$ & 1   & $2.14^{+0.05}_{-0.09}$ & M  & FVX      & cont. in F+V; H$\alpha$, [O{\sc iii}]5007 in X-NIR; merging with 067.2 \\
ALESS 067.2            &  53.179253         & -27.920749          & 2.1230 & 3   & $2.05^{+0.15}_{-0.13}$ & M  & X & BLANK but likely merging with 067.1 \\
ALESS 068.1            &  53.138888         & -27.653770          & ...      & 4   &  - & M  & VX       & BLANK \\
ALESS 069.1            &  52.890731         & -27.992345          & 4.2071 & 3   & $2.34^{+0.27}_{-0.44}$ & M  & D         & single line, poss. Ly$\alpha$ with asymmetric profile \\ 
ALESS 069.2            &  52.892226         & -27.991361          & ...      & 4   & - &  M  & M         & BLANK \\
ALESS 069.3            &  52.891524         & -27.993990          & ...      & 4   & - & M  & DM      & BLANK \\
ALESS 070.1            &  52.933425         & -27.643200          & 2.0918 & 3   & $2.28^{+0.05}_{-0.06}$ & M  & FX        & strong cont. in F; poss. Ly$\alpha$ in X-UVB \\ 
ALESS 071.1            &  53.273528         & -27.557831          & 3.6967 & 2   & $2.48^{+0.21}_{-0.11}$ & M  & V          & Ly$\alpha$ ($z=3.7006$); very bright line; N{\sc v} em ($z=3.6927$) \\ 
ALESS 072.1            &  53.168322         & -27.632807          & ...      & 4   & - & M  & DX       & poss. faint lines, no cont. \\			
ALESS 073.1            &  53.122046         & -27.938807          & 4.7649$^d$ & 1   & $5.18^{+0.43}_{-0.45}$ & M  & DF       & very broad Ly$\alpha$ and N{\sc v} em in D+F;  Ly$\alpha$ ($z=4.7648$), \\
                       &                    &                     & & & & & & N{\sc v} ($z=4.7649$) \\ 
ALESS 074.1            &  53.288112         & -27.804774          & ...      & 4   & $1.80^{+0.13}_{-0.13}$ & M  & DFV     & BLANK \\
ALESS 075.1            &  52.863303         & -27.930928          & 2.5450 & 1   & $2.39^{+0.08}_{-0.06}$ & M  & FVX & very interesting source; strong cont. in V+F;  [O{\sc iii}]4959 ($z=2.5452$), \\
                       &                    &                     & & & & & & [O{\sc iii}]5007 ($z=2.5447$) broad red components to [O{\sc iii}], H$\beta$ ($z=2.5451$),\\
                       &                    &                     & & & & & &  [O{\sc ii}] doublet ($z=2.5446$), H$\alpha$ ($z=2.5452$), Ly$\alpha$ in X ($z=2.5440$) \\
{\it ALESS 075.2}      &  {\it 52.865276}   & {\it -27.933116}    & {\it 2.2944}  & {\it 2}   & {\it 0.39$^{+0.02}_{-0.03}$} & {\it S}   & {\it DM} & {\it H$\alpha$, [N{\sc ii}] ($z=2.2941$), [S{\sc ii}] ($z=2.2886$) in M-$K$} \\
ALESS 075.4            &  52.860715         & -27.932144          & ...       & 4   & $2.10^{+0.29}_{-0.34}$ & M  & DM      & BLANK \\	
ALESS 076.1            &  53.384731         & -27.998786          & 3.3895  & 2   & - &  M  & DFMV & [O{\sc iii}]5007 + [O{\sc iii}]4959 in M; poss. Ly$\alpha$ ($z\sim3.3984$) in V\\
ALESS 079.1            &  53.088064         & -27.940830          & ...       & 4   & $2.04^{+0.63}_{-0.31}$ & M  & D         & BLANK \\	    	   		
ALESS 079.2            &  53.090004         & -27.939988          & 1.7693  & 1   & $1.55^{+0.11}_{-0.18}$ & M  & FVX & Strong H$\alpha$, [N{\sc ii}]6548, 6583 in X-NIR; structured lines- 2 components \\
ALESS 079.4            &  53.088261         & -27.941808          & ...       & 4   & - & M  & D         & BLANK \\	   
ALESS 080.1            &  52.928347         & -27.810244          & 4.6649  & 3   & $1.96^{+0.16}_{-0.14}$ & M  & FV        & poss Ly$\alpha$ in F \\
ALESS 080.2            &  52.927570         & -27.811376          & ...       & 4   & $1.37^{+0.17}_{-0.08}$ & M  & D         & BLANK \\
{\it ALESS 080.5}      &  {\it 52.923654}   & {\it -27.806318}    & {\it   1.3078}  & {\it 3}   & - & {\it S}   & {\it D} & {\it tentative [O{\sc ii}] + [Ne{\sc iii}]} \\
{\it ALESS 081.1}      &  {\it 52.864805}   & {\it -27.744336}    & {\it   ...}       & {\it 4}   & {\it 1.70$^{+0.29}_{-0.20}$} & {\it S}   & {\it V}          & {\it BLANK} \\
ALESS 082.1            &  53.224989         & -27.637470          & ...       & 4   & $2.10^{+3.27}_{-0.44}$ & M  & DFV      & BLANK \\
ALESS 084.1            &  52.977090         & -27.851568          & 3.9651  & 3   & $1.92^{+0.09}_{-0.07}$ & M  & DFM     & Ly$\alpha$ ($z=3.9639$), N{\sc v} ($z=3.9672$) in F;  cont. in F \\
ALESS 084.2            &  52.974388         & -27.851207          & ...       & 4   & $1.75^{+0.08}_{-0.19}$ & M  & DF        & cont. in F; poss faint lines	 \\
ALESS 087.1            &  53.212016         & -27.528187          & 2.3086  & 1   & $3.20^{+0.08}_{-0.47}$ & M  & FV & Ly$\alpha$ em ($z=2.3188$), Si{\sc iv} abs ($z=2.3050$), \\
                       &                    &                     & & & & & & Si{\sc ii} abs ($z=2.3019$) in V; Ly$\alpha$ offset from cont. \\
ALESS 088.1            &  52.978175         & -27.894858          & 1.2679  & 1   & $1.84^{+0.12}_{-0.11}$ & M  & FVMX & [O{\sc ii}]  ($z=1.2679$); [O{\sc ii}]3726,3729 visible in X-VIS \\ 
ALESS 088.2            &  52.980797         & -27.894529          & 2.5192  & 3   & - & M  & DM & C{\sc ii}]2326 em ($z=2.5227$), C{\sc iv} em ($z=2.5156$) in D \\ 
ALESS 088.5            &  52.982524         & -27.896446          & 2.2941  & 2   & $2.30^{+0.11}_{-0.50}$ & M  & DFV & strong cont. in V, poss break; Ly$\alpha$ em ($z=2.3021$), \\
                       &                    &                     & & & & & & He{\sc ii} ($z=2.2941$) in V \\
ALESS 088.11           &  52.978949         & -27.893785          & 2.3583  & 3   & $2.57^{+0.04}_{-0.12}$ & M  & D	       & C{\sc iii}] em ($z=2.3585$), Ly$\alpha$ em ($z=2.3581$) + break \\   
{\it ALESS 089.1}      &  {\it 53.202879}   & {\it -28.006079}    & {\it 0.6830}  & {\it 3}   & {\it 1.17$^{+0.06}_{-0.15}$} & {\it S}   & {\it F}           & {\it bright [O{\sc ii}] + cont} \\ 
\hline
    \end{tabular}
\end{sidewaystable}

\clearpage

\begin{sidewaystable}[h]
  \scriptsize
    \begin{tabular}{lllllllll}
    \multicolumn{9}{c}%
                {\textit{Continued from previous page}}\\
                \hline\hline
ALESS ID               & RA                 & DEC                 & z$_{spec}$    & Q$_{spec}$    & z$_{\rm phot}^a$            & M/S$^b$     & Instruments$^c$ & Notes  \\
                       & (J2000)            & (J2000)             &              &             &                           &             &             & \\ 
\hline
ALESS 094.1            &  53.281640         & -27.968281          & ...       &  4  & $2.87^{+0.37}_{-0.64}$ & M  & DV        & BLANK \\
ALESS 098.1            &  52.874654         & -27.956317          & 1.3745$^d$  &  1  & $1.63^{+0.17}_{-0.09}$ & M  & DFMVX  & [O{\sc ii}] ($z=1.3745$) brightest in F; cont. in M and F, \\
                       &                    &                     & & & & & & real H$\alpha$ under sky in X-NIR \\
ALESS 099.1            &  53.215910         & -27.925996          & ...       & 4   & - & M  & D          & BLANK \\	 
{\it ALESS 101.1}      & {\it  52.964987}   & {\it -27.764718}    & {\it 2.7999}  & {\it 2}   & {\it 3.49$^{+03.52}_{-0.88}$} &{\it S}   & {\it V} 	        & {\it Ly$\alpha$} \\
ALESS 102.1            &  53.398333         & -27.673061          & 2.2960  & 3   & $1.76^{+0.16}_{-0.18}$ & M  & FV & cont. in V, Ly$\alpha$ ($z=2.2931$), C{\sc iii}] ($z=2.2960$) in V \\
{\it ALESS 106.1}      &  {\it 52.915187}   & {\it -27.944236}    & {\it ...}       & {\it 4}   & {\it $7.00^{+0.00}_{-4.07}$} & {\it S}   & {\it DM}       & {\it BLANK} \\	
ALESS 107.1            &  52.877082         & -27.863647          & 2.9965  & 3   & $3.75^{+0.09}_{-0.08}$ & M  & VM & Ly$\alpha$ em ($z=2.9757$),  C{\sc iv} em ($z=2.9965$) in V; cont. in V+M; \\
                       &                    &                     & & & & &  & poss. [O{\sc ii}], [O{\sc iii}] in M \\
ALESS 107.3            &  52.878013         & -27.865465          & ...       & 4   & $2.12^{+1.54}_{-0.81}$ & M  & D	       &  BLANK \\
ALESS 110.1            &  52.844411         & -27.904784          & ...       & 4   & $2.55^{+0.70}_{-0.50}$ & M  & FMV      & BLANK \\
ALESS 110.5            &  52.845677         & -27.904005          & ...       & 4   & - & M  & DM      & BLANK \\ 
ALESS 112.1            &  53.203596         & -27.520362          & 2.3154  & 1   & $1.95^{+0.15}_{-0.26}$ & M  & FGV   & Ly$\alpha$ em ($z=2.3122$) + cont. in V , H$\alpha$ ($z=2.3145$), \\
                       &                    &                     & & & &  & & poss [O{\sc iii}]5007 ($z=2.3157$), H$\beta$ em ($z=2.3160$) in G \\		
ALESS 114.2            &  52.962945         & -27.743693          & 1.6070  & 1   & $1.56^{+0.07}_{-0.07}$ & M  & FV       & strong cont in F+V, [O{\sc ii}] doublet in F ($z=1.6070$) \\ 
ALESS 115.1            &  53.457070         & -27.709609          & 3.3631  & 3   & - & M  & V         & cont., poss Ly$\alpha$ em ($z=3.3631$) \\
ALESS 116.1            &  52.976342         & -27.758039          & ...       & 4   & $3.54^{+1.47}_{-0.87}$ & M  & FV       & BLANK \\				
ALESS 116.2            &  52.976826         & -27.758735          & ...      & 4   & $4.02^{+1.19}_{-2.19}$ & M  & F          & BLANK \\			
ALESS 118.1            &  52.841347         & -27.828161          & 2.3984  & 3   & $2.26^{+0.50}_{-0.23}$ & M  & DFV & strong cont in F+V, Ly$\alpha$ abs + break, C{\sc iv} em ($z=2.3984$) in V  \\
ALESS 119.1            &  53.235993         & -28.056988          & ...       & 4   & $3.50^{+0.95}_{-0.35}$ & M  & V         & BLANK \\
ALESS 122.1            &  52.914768         & -27.688792          & 2.0232$^d$  & 1   & $2.06^{+0.05}_{-0.06}$ & M & FV        & very strong blue cont. and abs. lines. V:  C{\sc ii}] abs ($z=2.0197$), \\
                       &                    &                     & & & & & & Si{\sc iv} abs ($z=2.0229$), He{\sc ii} em ($z=2.0282$), \\
                       &                    &                     & & & & & & Very broad C{\sc iv} and Si{\sc ii} blended abs.; C{\sc iii}] ($z=2.0222$). \\
                       &                    &                     & & & & & & F:  Fe{\sc ii}\,2344, Fe{\sc ii}\,2375, Fe{\sc ii}\,2383 abs \\
ALESS 124.1            &  53.016843         & -27.601769          & ...       & 4   & $6.07^{+0.94}_{-1.16}$ & M & FV         & poss faint lines \\
ALESS 126.1            &  53.040033         & -27.685466          & ...       & 4   & $1.82^{+0.28}_{-0.08}$ & M  & V	      & BLANK \\
\hline
    \end{tabular} \caption{Notes: The 22 ALESS SMGs not targeted in
 our spectroscopy programme (and without redshifts from literature)
 are not listed here. The {\sc supp} SMGs are shown in
 italics. $z_{spec}=-99$ means we could not determine a spectroscopic
 redshift. $^a$Photometric redshifts from S14. Those SMGs without a
 photometric redshift have poor photometric constraints (detections in
 $<4$ bands). $^b$M\,=\,{\sc main} catalog, S\,=\,{\sc supp}
 catalog. $^c$F\,=\,VLT/FORS2, V\,=\,VLT/VIMOS, X\,=\,VLT/XSHOOTER,
 M\,=\,Keck/MOSFIRE (Band $H$ or $K$), D\,=\,Keck/DEIMOS,
 G\,=\,Gemini/GNIRS. $^d$These redshifts are for the six sources which
 also have literature spectroscopic redshifts described
 in \S~\ref{sec:anal2}. The quality flag (Q) for the spectroscopic
 redshifts is Q\,=\,1 for secure redshifts; Q\,=\,2 for redshifts
 measured from only one or two strong lines; Q\,=\,3 for tentative
 redshifts measured based on one or two very faint features; Q\,=\,4
 for those sources which were targeted but no redshift could be
 determined.}
\end{sidewaystable}

\label{appendix:smg}

\clearpage

\section[]{Ancillary Redshifts}
\label{sec:infill}
When designing the slit masks, we in-filled the unused portions masks
(not targeting the high-prioroty SMGs) with other candidate
high-redshift galaxies, in particular with mid-, far-infrared or radio
selected galaxies.  Here, we provide the details of the galaxies
targeted.

The ID for each galaxy relates to the input catalogue from which a
target was selected.  These are summarised as:

\noindent{\it 101--500:} Statistically Robust or Tentative candidate
LESS SMG multiwavelength counterparts from \citet{Biggs11} (see also
\citealt{Wardlow11}) but which were later shown by ALMA observations
to be incorrect IDs \citep{Hodge13}.\\

\noindent{\it 500--700:} Robust or tentative IDs for LESS sources with
signal-to-noise of SNR\,=\,2.7--3.7$\sigma$ in the original LESS map.
These IDs for ``faint SMGs'' are derived using 1.4\,GHz radio emission
\citep{Biggs11} but have not yet been confirmed (or ruled out) by
ALMA.  \\

\noindent{\it 700--1000:} Galaxies in the LESS submillimetre error circles which have
photometric redshifts that are consistent with the ALESS photometric
redshifts \citep{Wardlow11}.\\

\noindent{\it 1000--3000:} 24+70$\mu$m-selected galaxies from the {\it
  Spitzer} FIDEL survey without pre-existing spectroscopic redshifts
\citep{Magnelli09}.\\

\noindent{\it 4000--4300:} {\it Chandra} X-ray sources from the 2\,Ms or 4\,Ms
surveys \citep[e.g.][]{Lehmer05,Luo08}.\\

\begin{figure*}
\begin{center}{
\psfig{file=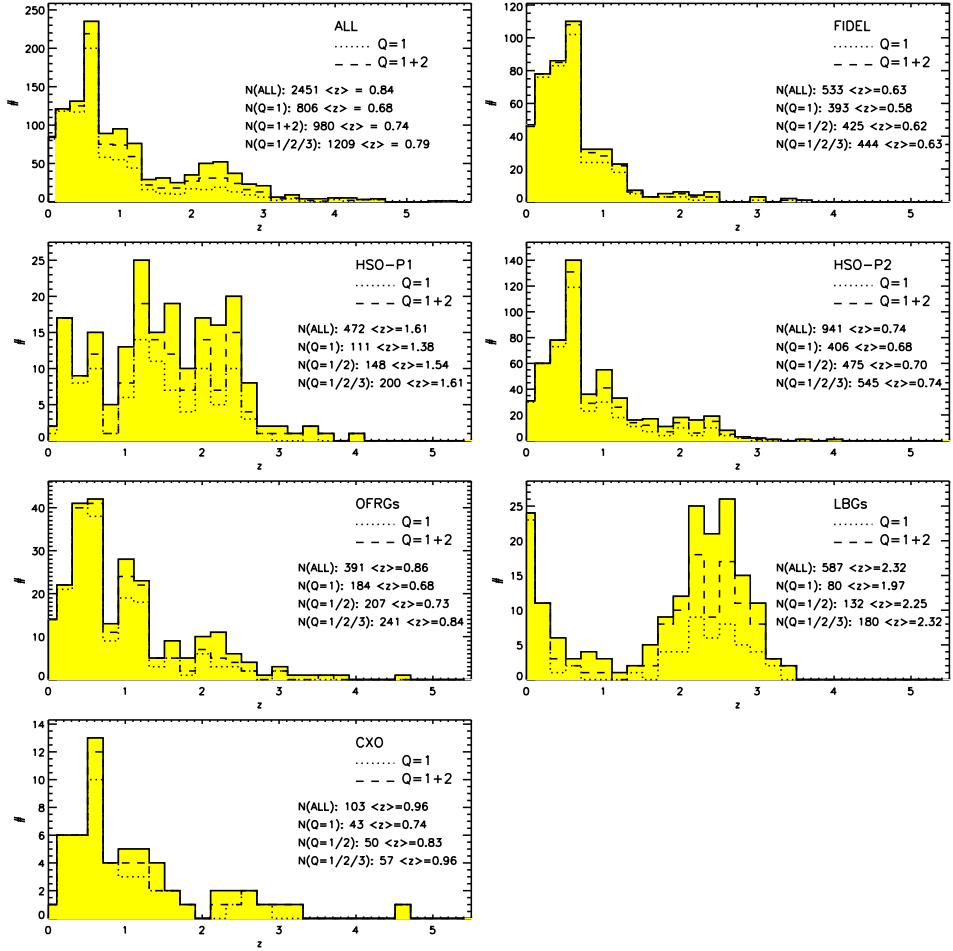,width=5in,angle=90}
\caption{Spectroscopic redshift distributions for the various galaxy
  population targeted during the spectroscopic campaign.  In each
  panel, we show the redshift distribution for all galaxies, but also
  show the histograms for the best quality (Q\,=\,1) spectra, and those
  with Q\,=\,1 \& 2.  The number of galaxies with spectroscopic redshifts
  (and the median redshift) are also given in the panels.  {\it Top
    Left}: Redshift distribution for {\sc ALL} galaxies targeted;
  {\it Top Right:} Redshift distribution for 24$\mu$m selected
  galaxies from the FIDEL survey; {\it Middle Left:} Redshift
  distribution for optically faint radio galaxies (OFRGs); {\it Middle
    Right:} Redshift distribution for the LBGs, BX/BMs and Ly$\alpha$
  emitters; {\it Bottom Left:} Redshift distribution for {\it Chandra}
  X-ray sources.}
\label{fig:NzALL}
}\end{center}
\end{figure*}

{\it 5000--6000:} Galaxies from the {\it Herschel}\,/\,SPIRE images
which peak at 350$\mu$m (and which have been identified and deblended
using the 24$\mu$m positions as priors; \citealt{Roseboom10}).
Individual redshifts for these sources will be published in Oliver et
al.\ (in prep), although we include the redshift distributio in
Fig.~\ref{fig:NzALL}.

{\it 6000-9800:} Galaxies from the {\it Herschel}\,/\,SPIRE images
which peak at 250$\mu$m or 350$\mu$m (and which have been identified
and deblended using the 24$\mu$m positions as priors;
\citealt{Roseboom10}.  Individual redshifts for these sources will be
published in Oliver et al.\ (in prep), although we include the
redshift distributio in Fig.~\ref{fig:NzALL}.

\noindent{\it 50000--51000:} Optically faint radio galaxies (OFRGs) from the
JVLA 1.4\,GHz survey of this field.  These radio sources are typically
brighter than $>$20\,$\mu$Jy at 1.4\,GHz but have optical magnitudes
fainter than $I_{\rm AB}$\,=\,22.\\

\noindent{\it 70000--72000:} Optically (colour) selected galaxies.  These
comprise a mix of $z\sim$\,2 Lyman$\alpha$ emitting galaxies, BM/BX
galaxies and Lyman break galaxies at $1.5<z<3.5$.\\

\noindent{\it 80000--89999:} Galaxies which were not in any of the
other prior catalogs but which could still be placed on the masks.\\

\noindent{\it 90000--90200:} $B$- or $V$-band drop-out galaxies (i.e.\ candidate
$z\gsim$\,2.5 or $z\gsim$\,3.5 galaxies).\\

Any source that is labelled with a {\it ``b''} suffix denotes a
secondary galaxy that happened to lie on the slit, but is not the
primary target.

We also note that the catalogs are not unique (a galaxy could be an
ALMA source that is also in the FIDEL 24$\mu$m catalog, a radio
catalog, a BX/BM and also a {\it Chandra} X-ray source).  In those instances, the
object will only appear once in the table, but under the ID from which
it was selected for slit placement (i.e.\ there are no RA\,/\,Dec
repeats).  As in Table~2, the instrument IDs are denoted by
F\,=\,VLT\,/\,FORS2, V\,=\,VLT\,/\,VIMOS, X\,=\,VLT\,/\,XSHOOTER,
M\,=\,Keck\,/\,MOSFIRE, D\,=\,Keck\,/,\,DEIMOS, and
G\,=\,Gemini\,/\,GNIRS.  The quality flag (Q) for the spectroscopic
redshifts is Q\,=\,1 for secure redshifts; Q\,=\,2 for redshifts
measured from only one or two strong lines; Q\,=\,3 for tentative
redshifts measured based on one or two very faint features; Q\,=\,4
for those sources which were targeted but no redshift could be
determined.  The redshift distribution for each of these sub-samples
is shown in Fig.~\ref{fig:NzALL}.

%
%

\clearpage

\label{appendix:all}
\begin{table}
  \begin{tabular}{llllllllllll}
    \multicolumn{9}{c}%
                {\textit{Table 3. Spectroscopic redshifts for the full sample}} \\
\hline\hline
ID           &    RA     &      DEC           &  z$_{\rm spec}$     &     Q &   Inst    & \vline \ \ ID           &    RA     &      DEC           &  z$_{\rm spec}$     &     Q &   Inst    \\ 
             & (J2000)   & (J2000)            &                     &       &                  & \vline \ \       & (J2000)   & (J2000)            &                     &       &           \\ 
\hline
101 & 53.30820 & -27.93445 & 4.6892 & 1 & F & \vline \ \ 104 & 53.26036 & -27.94606 & 1.9469 & 3 & VF\\
106 & 52.90094 & -27.91398 & 2.3484 & 3 & VMF & \vline \ \ 107 & 52.89957 & -27.91209 &  ...   & 4 & VMF\\
108 & 52.89780 & -27.90952 &  ...   & 4 & VF & \vline \ \ 109 & 52.90089 & -27.91278 & 3.0159 & 2 & V\\
110 & 52.87580 & -27.98573 & 1.4135 & 1 & F & \vline \ \ 112 & 52.87865 & -27.98229 & 0.4342 & 1 & F\\
113 & 53.23814 & -28.01708 & 1.3648 & 3 & VF & \vline \ \ 114 & 53.23651 & -28.01645 &  ...   & 4 & VF\\
116 & 53.31593 & -27.76045 & 0.7516 & 1 & VF & \vline \ \ 117 & 53.02072 & -27.51948 & 0.9610 & 2 & VF\\
118 & 53.01840 & -27.52046 & 0.7283 & 3 & VF & \vline \ \ 119 & 53.04730 & -27.87038 &  ...   & 4 & F\\
280 & 53.08039 & -27.87200 &  ...   & 3 & V & \vline \ \ 122 & 53.19980 & -27.90448 & 3.1977 & 3 & V\\
123 & 53.20365 & -27.71445 &  ...   & 4 & VF & \vline \ \ 123b & 53.20339 & -27.71603 & 2.8382 & 2 & V\\
124 & 52.96913 & -28.05492 &  ...   & 4 & V & \vline \ \ 127 & 53.07793 & -27.62877 &  ...   & 4 & V\\
131 & 53.03317 & -27.97311 & 0.9607 & 3 & VF & \vline \ \ 133 & 53.37387 & -27.57901 & 1.2382 & 2 & V\\
\hline
\label{tab:all_z}
\end{tabular}
\begin{flushleft}
\caption{{\sc Notes}: Only a subsample of the table is shown here for
  form and content.  The full table is available in the online version
  of the article.  The labels in the instrument column are defined as:
  F\,=\,VLT\,/\,FORS2, V\,=\,VLT\,/\,VIMOS, X\,=\,VLT\,/\,XSHOOTER,
  M\,=\,Keck\,/\,MOSFIRE (Band $H$ or $K$), D\,=\,Keck\,/\,DEIMOS,
  G\,=\,Gemini\,/\,GNIRS.  The quality flag (Q) for the spectroscopic
  redshifts is Q\,=\,1 for secure redshifts; Q\,=\,2 for redshifts
  measured from only one or two strong lines; Q\,=\,3 for tentative
  redshifts measured based on one or two very faint features; Q\,=\,4
  for those sources which were targeted but no redshift could be
  determined.}
\end{flushleft}
\end{table}

\end{document}